\documentclass[fleqn,usenatbib]{mnras}

\usepackage{newtxtext,newtxmath}
\usepackage{soul}

\usepackage[T1]{fontenc}
\usepackage{multirow}
\usepackage[table]{xcolor}

\DeclareRobustCommand{\VAN}[3]{#2}
\let\VANthebibliography\thebibliography
\def\thebibliography{\DeclareRobustCommand{\VAN}[3]{##3}\VANthebibliography}

\usepackage{graphicx}	
\usepackage{amsmath}	
\usepackage{float}

\title[Pulsars glitch monitoring at IAR]{First results of the glitching pulsars monitoring program at the Argentine Institute of Radioastronomy}

\author[Zubieta et al.]
{
\newauthor{
Ezequiel Zubieta$^{1}$\thanks{Fellow of CONICET, Argentina. E-mail: ezubieta@iar.unlp.edu.ar},
Ryan Missel$^{2}$,
Valentina Sosa Fiscella$^{1,4}$,
Carlos O. Lousto$^{3,4}$\thanks{E-mail: colsma@rit.edu (COL)},}
\newauthor{
Santiago del Palacio$^{1,6}$,
Federico G. L\'opez Armengol$^{4}$,
Federico Garc\'{i}a$^{1,5}$,
Jorge A. Combi$^{1,5,7}$,}
\newauthor{
Linwei Wang$^{2}$,
Luciano Combi$^{1,4}$,
Guillermo Gancio$^{1}$, 
Carolina Negrelli$^{5}$, 
Eduardo M. Guti\'errez$^{1}$
} 
\\
\\
$^{1}$Instituto Argentino de Radioastronom\'ia (CCT La Plata, CONICET; CICPBA; UNLP), C.C.5, (1894) Villa Elisa, Buenos Aires, Argentina.\\
$^{2}$Golisano College of Computing and Information Sciences, Rochester Institute of Technology Rochester, NY 14623, USA\\
$^{3}$School of Mathematical Sciences, Sciences Rochester Institute of Technology Rochester, NY 14623, USA\\
$^{4}$Center for Computational Relativity and Gravitation, Rochester Institute of Technology, 85 Lomb Memorial Drive, Rochester,New York 14623, USA\\
$^{5}$Facultad de Ciencias Astron\'omicas y Geof\'{\i}sicas, Universidad Nacional de La Plata, Paseo del Bosque, B1900FWA La Plata, Argentina\\
$^{6}$Department of Space, Earth and Environment, Chalmers University of Technology, SE-412 96 Gothenburg, Sweden\\
$^{7}$Departamento de F\'\i sica (EPS), Universidad de Ja\'en, Campus Las Lagunillas s/n, A3, 23071 Ja\'en, Spain\\
}

\date{Accepted XXX. Received YYY; in original form ZZZ}

\pubyear{2022}

\graphicspath{{./}{figures/}}

\newcommand{\NOTE}[1]{{\bf \color{red} #1 }}

\begin{document}
\label{firstpage}
\pagerange{\pageref{firstpage}--\pageref{lastpage}}
\maketitle

\begin{abstract}
We report here on the first results of a systematic monitoring of southern glitching pulsars at the Argentine Institute of Radioastronomy that started in the year 2019. We detected a major glitch in the Vela pulsar (PSR J0835$-$4510) and two small-glitches in PSR J1048$-$5832. For each glitch, we present the measurement of glitch parameters by fitting timing residuals.
We then make an individual pulses study of Vela in observations before and after the glitch. We selected 6 days of observations around the major glitch on 2021 July 22 and study their statistical properties with machine learning techniques. We use Variational AutoEncoder (VAE) reconstruction of the pulses to separate them clearly from the noise. We perform a study with Self-Organizing Maps (SOM) clustering techniques to search for unusual behavior of the clusters during the days around the glitch not finding notable qualitative changes.
We have also detected and confirm recent glitches in PSR J0742$-$2822 and PSR J1740$-$3015.
\end{abstract}

\begin{keywords}
pulsars: Vela -- methods: observational -- methods: statistical
\end{keywords}



\section{Introduction}\label{sec:intro}

Pulsars are a sub-type of neutron stars that present pulsed emission, predominantly in the radio band.
The very high moment of inertia of the neutron stars renders them with an extraordinarily stable rotation, making pulsars one of the most accurate clocks in the Universe.
Although pulsars have extremely stable periods over time, some young pulsars are prone to have glitches: sudden changes in their period due to changes in the interior of the star. Discovered 50~years ago, nowadays almost 200~pulsars are known to glitch \citep{Manchester:2018jhy}. Southern \citep{Yu2013} and northern \citep{Espinoza:2011pq,Fuentes:2017bjx} based surveys provide comprehensive catalogs such as ATNF and JBO \footnote{\url{http://www.atnf.csiro.au/people/pulsar/psrcat/glitchTbl.html}\\ \url{http://www.jb.man.ac.uk/pulsar/glitches/gTable.html}}. The physical mechanism behind these glitches is still not well understood.

The Vela Pulsar (PSR B0833$-$45 / PSR J0835$-$4510) is one of the most active pulsars in terms of glitching, counting 21 in the last 50+ years. Although erratic, this pulsar exhibits major glitches every 2--3 years.
On the theoretical modeling, superfluidity is required \citep{2018ApJ...865...23G}, as the rotational dynamics of the neutron superfluid that resides under the outer crust (or surface) are necessary to explain the large Vela
glitches \citep{2012PhRvL.109x1103A,2015IJMPD..2430008H}. The glitch magnitude gives some idea of the available angular momentum reservoir, which in turn gives us information about the moment of inertia of the superfluid that produces such glitches. For a recent study of the 2016 pulse-to-pulse glitch in the Vela pulsar
and its use to estimate of the superfluid moments of inertia, see \cite{2020A&A...642A.223M}.
Observations can also be used to estimate the mass of the neutron stars \citep{2015SciA....1E0578H,2020MNRAS.492.4837M,khomenko_haskell_2018} and the post-glitch relaxation properties should provide a handle on the so-called mutual friction \citep[involving neutron superfluid
vortices and their mutual friction is related to their interaction
with other stellar components such as crust and core;][]{2018ApJ...865...23G}. Moreover, a detailed study of the pulsed emission can provide further insight on the physics of glitches \citep{2020ApJ...897..173B}. In particular, the analysis of the single pulses in the 2016 Vela glitch showed an atypical behaviour of a few pulses around the glitch, including a null, namely no pulse at all seen, which  revealed that the glitch also affects the pulsar  magnetosphere \citep{2018Natur.556..219P}. Unfortunately, the unpredictable character of the glitch phenomenon makes it extremely difficult to observe. A valid question is whether it is possible that information of a glitch precursor exists before the glitch event itself, and also if we can learn more from observations during the relaxation phase just after the glitch.  

Since 2019, the Pulsar Monitoring in Argentina\footnote{\url{https://puma.iar.unlp.edu.ar}} (PuMA) collaboration has been monitoring with high cadence a set of pulsars from the southern hemisphere that had shown glitches before \citep{Gancio2020}. The observations are carried out with the antennas from the Argentine Institute of Radio astronomy (IAR). 
A major goal of our observing campaign is the close follow-up of the Vela pulsar. The consistency of our monitoring allowed us to detect its last two large glitches: the one on 2019 February 1st  \citep{atel_vela} was measured with observations three days before and three days after the event, while the one on 2021 July 22nd was observed just one hour after the glitch, and we first reported it in \cite{2021ATel14806....1S}. We plan to continue monitoring the Vela pulsar to attempt to capture a future glitch "live" during our 3.5-h daily observations.

Moreover, as the Vela pulsar is very bright, we are able to detect its individual (single) pulses. Recently, in \cite{Lousto:2021dia} we performed an individual-pulses study of a sample of our daily observations that span over three hours (around 120,000 pulses per observation). 
We selected 4 days of observations in January--March 2021 
and studied their statistical properties with machine learning techniques. We first used density based DBSCAN clustering techniques, associating pulses mainly by amplitudes, and found a correlation between higher amplitudes and earlier arrival times. We also found a weaker (polarization dependent) correlation with the mean width of the pulses. We identified clusters of the so-called mini-giant pulses, with $\sim10$ times the average pulse amplitude. We then performed an independent study, using the Variational AutoEncoder (VAE) reconstruction \citep{kingma2014autoencoding} of the pulses to separate them clearly from the noise and select one of the days of observation to train VAE and apply it to the rest of the observations. We applied to those reconstructed pulses Self-Organizing Maps (SOM) clustering techniques \citep{teuvo1988som} to determine 4 clusters of pulses per day per radio telescope and concluded that our main results were robust and self-consistent. These results supported models for emitting regions at different heights (separated each by roughly a hundred km) in the pulsar magnetosphere.
Given the success of these techniques we apply them here on the major glitch event on 2021 July 22nd, for which we have collected data daily around that glitch. 

The goals of our observing campaign also include the creation of updated ephemeris of glitching pulsars that can be relevant for other studies, such as the search of continuous gravitational waves detectors such as LIGO. 
In addition to Vela, we are currently monitoring the pulsars mentioned in \cite{Gancio2020}, PSR J0738$-$4042, J0742$-$2822, J1048$-$5832 J1430$-$6623, J1644$-$4559, J1709$-$4429, J1721$-$3532, J1731$-$4744, J1740$-$3015, 
and plan to extend the list to other accessible (bright) glitching pulsars. 
In this work we present our observations of the pulsars J0835$-$4510 and J1048$-$5832 and provide a detailed analysis of their most recent glitches.
We find a large Vela glitch on 2021 July 22nd and two mini-glitches (the lowest amplitude so far from the previous 7 glitches recorded) on 2020 December 20th and on 2021 November 20th.

\section{Pulsars Glitch Monitoring Program at IAR}\label{sec:Pugliese}


The IAR observatory is located near the city of La Plata, Argentina (local time UTC$-3$), at latitude $-34\degr 51\arcmin 57\arcsec.35$ and longitude $58\degr 08\arcmin 25\arcsec.04$. 
It has two 30~m single-dish antennas, A1 and A2, aligned on a North--South direction and separated by $120$~m. These radio telescopes cover a declination range of $-90\degr < \delta <-10\degr$ and an hour angle range of two hours east/west, $-2\,\mathrm{h} <t< 2\, \mathrm{h}$. Although the IAR is not located in a radio frequency interferences (RFI) quiet zone, the analysis of the RFI environment presented in \cite{Gancio2020} showed that the radio band from 1~GHz to 2~GHz has a low level of RFI activity that is suitable for radio astronomy.




Major upgrades have been done to both antennas since 2014. 
Some of these include the installation of two digitizer boards of 56~MHz bandwidth that can be used as consecutive bands to give a total 112-MHz bandwidth on a single polarization. We note that the receiver in A2 is different from the one in A1, having fewer radio frequency components and larger RF bandwidth, which translates in different responses for each antenna. A detailed description of the characteristics of the current front end in A1 and A2 are given in 
\cite{Gancio2020}. We highlight that a major asset of IAR's observatory is its availability for high-cadence long-term monitoring of bright sources.  

We are carrying out an intensive monitoring campaign of known bright glitching pulsars in the southern hemisphere in the L-band (1400~MHz) using the two IAR antennas. 
Our observational program includes high-cadence observations (up to daily) with a duration of up to 3.5~h per day. This builds a unique database aimed to detect and characterise both large and small (mini-) glitches. In addition, the intensive monitoring also gives a significant chance that a glitch could be observed "live", a goal that has been achieved only on extremely rare occasions by other monitoring programs  \citep[e.g.][]{2018Natur.556..219P,Flanagan1990RapidRO,2002ApJ...564L..85D}.

For both antennas, the data is acquired with a timing resolution of $146~\mu$s. In the case of A1, we use 128 channels of 0.875~MHz centered at 1400 MHz in single (circular) polarization mode, whereas for A2 we use 64 channels of 1~MHz centered at 1416 MHz and in dual polarization (both circular polarizations added). When possible, we observe each target with both antennas independently, in order to control systematic effects. Unfortunately, a clock issue affected the observations with A2 during the period MJD 59400--59435 (July 5th to August 9th, 2022), which thus had to be excluded in the timing analysis of the residuals.

Here we analyse close to $270$~h of data of Vela J0835$-$4510 ($145.6$~h with A1, $122.7$~h with A2) taken in the period (MJD) 59371--59463. These observations include an almost daily monitoring close to the 2021 glitch (Sect.~\ref{sec:J0835}). In addition, we also study $730$~h of data of J1048$-$5832 ($553.7$ h with A1, $177.3$ h with A2) during the period (MJD) 59031--59729. We note that, when possible, the observations of the Vela pulsar span for the maximum tracking range of the antennas, which is $\sim 3.5$~h, while for J1048$-$5832 they last $<2.5$~h, due to an overlap with Vela (which is prioritized in our schedule).

\section{Glitches: analysis and results}

Pulsar rotation can be monitored by observing the times of arrival (ToAs) of their pulses. To extract information from the ToAs, one introduces a timing model that is essentially a mathematical model aimed to predict the ToAs. The difference between the predicted and observed ToAs can reveal the limitations of the timing model to represent the pulsar behaviour, which can be used to derive information of the pulsar itself. 

In the timing model, the temporal evolution of the pulsar phase is modeled as a Taylor expansion \citep{Basu},
\begin{equation}\label{eq:timing-model}
    \phi(t)=\phi+\nu(t-t_0)+\frac{1}{2}\dot{\nu}(t-t_0)^2+\frac{1}{6}\Ddot{\nu}(t-t_0)^3,
\end{equation}
where $\nu$, $\dot\nu$ and $\ddot\nu$ are the rotation frequency of the pulsar, and its first and second derivatives.

When a glitch occurs, the pulsar suffers a sudden jump in its rotation frequency. This spin up can be introduced in the timing model as a change in the phase of the pulsar modeled as \citep{glitch-timing}
\begin{multline}\label{eq:glitch-model}
    \phi_\mathrm{g}(t) = \Delta \phi + \Delta \nu_\mathrm{p} (t-t_\mathrm{g}) + \frac{1}{2} \Delta \dot{\nu}_\mathrm{p} (t-t_\mathrm{g})^2 + \\  \frac{1}{6} \Delta \Ddot{\nu}(t-t_\mathrm{g})^3+
    \left[1-\exp{\left(-\frac{t-t_\mathrm{g}}{\tau_\mathrm{d}}\right)} \right]\Delta \nu_\mathrm{d} \, \tau_\mathrm{d},
\end{multline}
where $\Delta \phi$ is the offset in pulsar phase, $t_\mathrm{g}$ is the glitch epoch, and $\Delta \nu_\mathrm{p}$, $\Delta \dot{\nu}_\mathrm{p}$ and $\Delta \Ddot{\nu}$ are the respective permanents jumps in $\nu$, $\dot\nu$ and $\Ddot{\nu}$ relative to the pre-glitch solution. Finally, $\Delta \nu_\mathrm{d}$ is the transient increment in the frequency that decays on a timescale $\tau_\mathrm{d}$. From these parameters one can calculate the degree of recovery, $Q$, which relates the transient and permanent jumps in frequency as $Q=\Delta \nu_\mathrm{d} / \Delta \nu_\mathrm{g}$. At last, two commonly used parameters in the literature are the instantaneous changes in the pulse frequency and its first derivative
(at the glitch epoch), which can be described as 
\begin{align}
    \Delta \nu_\mathrm{g} &= \Delta \nu_\mathrm{p} + \Delta \nu_\mathrm{d} \\ 
    \Delta \dot{\nu}_\mathrm{g} &= \Delta \dot{\nu}_\mathrm{p} - \frac{\Delta \nu_\mathrm{d}}{\tau_\mathrm{d}} \, .
\end{align} 

The initial sets of parameters for the timing models were retrieved from the ATNF pulsar catalogue \citep{ManchesterATNF2005}, and then updated by ourselves. 
For the data reduction we used the software \texttt{PRESTO}  \citep{2003ApJ...589..911R, 2011ascl.soft07017R}. In particular, we used the tasks \texttt{rficlean} to remove RFIs and \texttt{prepfold} for folding the observations. The ToAs were subsequently determined from the folded observations using the Fourier phase gradient-matching template fitting \citep{1992PTRSL.341..117T} implemented in the \texttt{pat} package in \texttt{psrchive} \citep{2004PASA...21..302H}. 
Given the similarities between A1 and A2, we used the same template for observations with either antenna without introducing additional error. The template was created by applying a smoothing wavelet method to the pulse profile of a high signal-to-noise observation not included in the posterior timing analysis. Finally, the timing residuals were calculated using the pulsar timing software package \texttt{Tempo2} \citep{Hobbs2006} in a Python interface provided by \texttt{libstempo}\footnote{\url{https://github.com/vallis/libstempo}.}.

\subsection{Mini-glitches detection in PSR J1048$-$5832}\label{sec:J1048} 

PSR J1048$-$5832 has a period $P=123~\mathrm{ms}$ and a period derivative $\dot{P}=9.61 \times 10^{-14}$~s~s$^{-1}$, which leads to a characteristic age $\tau_c=P/2\dot{P} \sim 20~\mathrm{kyr}$.
In 2009, \textit{Fermi}-LAT detected its gamma-ray pulsations (photon energies $>0.1$~GeV), adding PSR~J1048$-$5832 to the list of young gamma-ray pulsars in the Galactic plane \citep{Fermi-LAT:2009puq}.
In addition, an optical counterpart has been searched (but not found) with deep VLT imaging by
\cite{Danilenko:2013gg}, and periodic amplitude modulation in PSR J1048$-$5832 interpreted as periodic mode-changing has been revealed with high-sensitivity radio observations by \cite{Yan:2019lti}. 

\begin{figure}
    \centering
    \includegraphics[width=\linewidth]{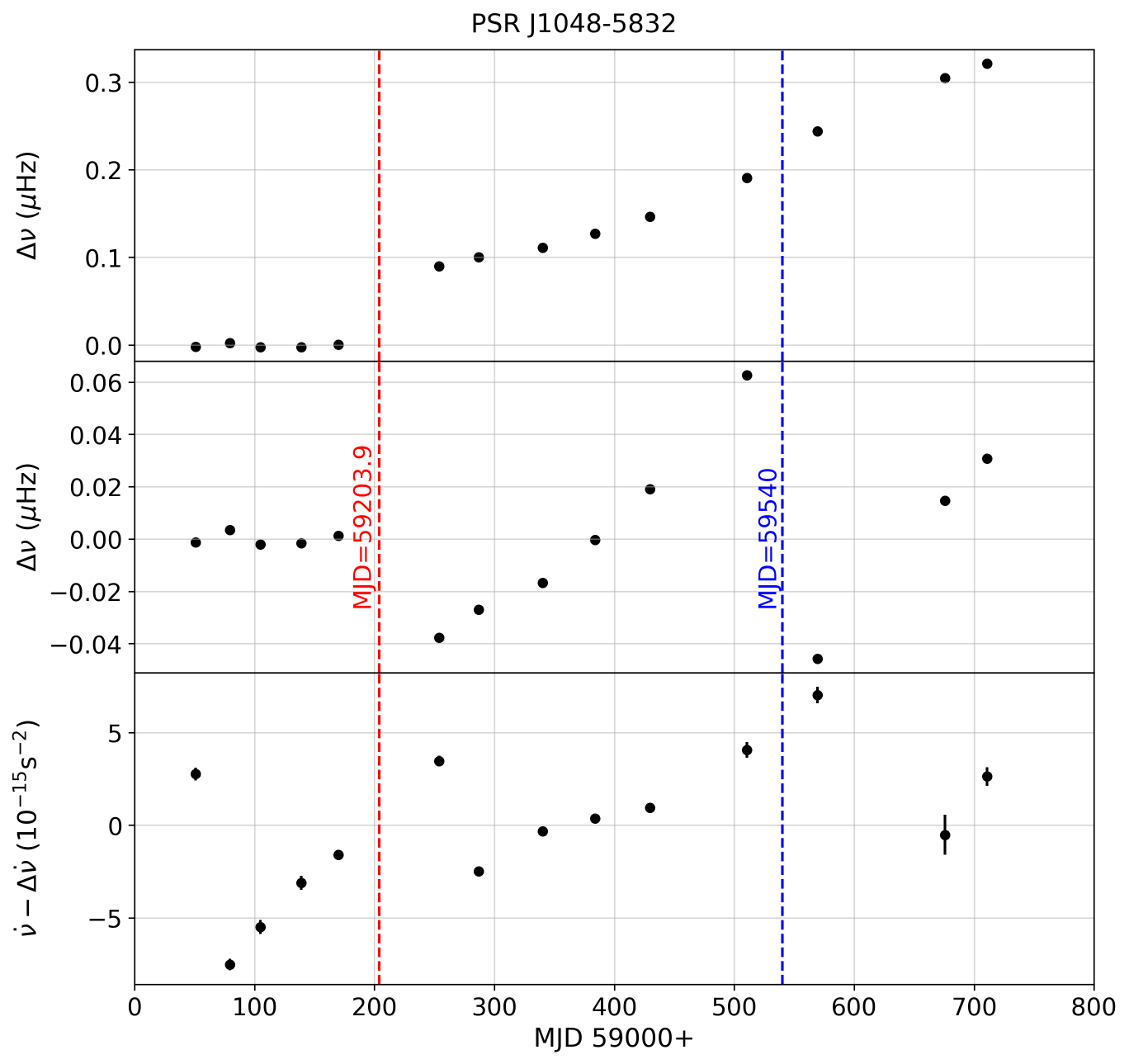}
    \caption{Timing analysis of PSR J1048$-$5832. \textit{Top:} variations in the rotational frequency $\Delta\nu$ relative to the solution before the first glitch. 
    \textit{Center:} expanded plot of $\Delta\nu$. Here the mean value of $\Delta\nu$ between the first and second glitch was subtracted from the data for that range of days, and the mean value of $\Delta\nu$ after the second glitch was subtracted from the data after that glitch. \textit{Bottom:} variations of the frequency first derivative $\Delta\dot{\nu}$ relative to the mean value of $\dot{\nu}$ along the whole data span. The vertical dashed lines mark the epochs of the two glitches.}
    \label{fig:1048}
\end{figure}

Seven glitches have been reported so far for this pulsar, observed between years 1992 and 2014. Here, we report the detection of two new glitches between 2020 and 2022, more precisely on MJD 59203.9(5) \citep{2022BAAA...63..262Z} and MJD 59540(2). We used the \texttt{glitch} plug-in in \texttt{tempo2} \citep{Hobbs2006} to subdivide the observations in blocks of 50--100 days and then fit $\nu_0$ and $\dot\nu_0$ in each of these blocks. The results are displayed in Fig.~\ref{fig:1048}. Our analysis reveals a frequency jump consistent with a glitch on MJD 59203.9, after which there is a continuous increase in the frequency relative to the pre-glitch model. This type of behaviour is unusual but it has also been observed in PSR J2219$-$4754 \citep{2022arXiv220612886Z} and PSR J0147$+$5922 \citep{2010MNRAS.404..289Y}.

The dataset before the first glitch covers the timespan MJD 59031--59204 and accounts for 71 observations with A1 and 47 observations with A2. In the timespan between the first and the second glitch, MJD 59205--59513, we have 57 observations with A1 and 17 observations with A2. Finally, for the epoch after the second glitch, our dataset covers the timespan MJD 59571--59730, in which we have 16 observations with A1 and 16 observations with A2. All observations are folded in radio-frequency with a fixed dispersion measure $DM=128.678(3)$~pc\,cm$^{-3}$ from the ATNF catalogue ({\url{http://www.atnf.csiro.au/people/pulsar/psrcat/}).

\begin{figure}
    \centering
    \includegraphics[width=\linewidth]{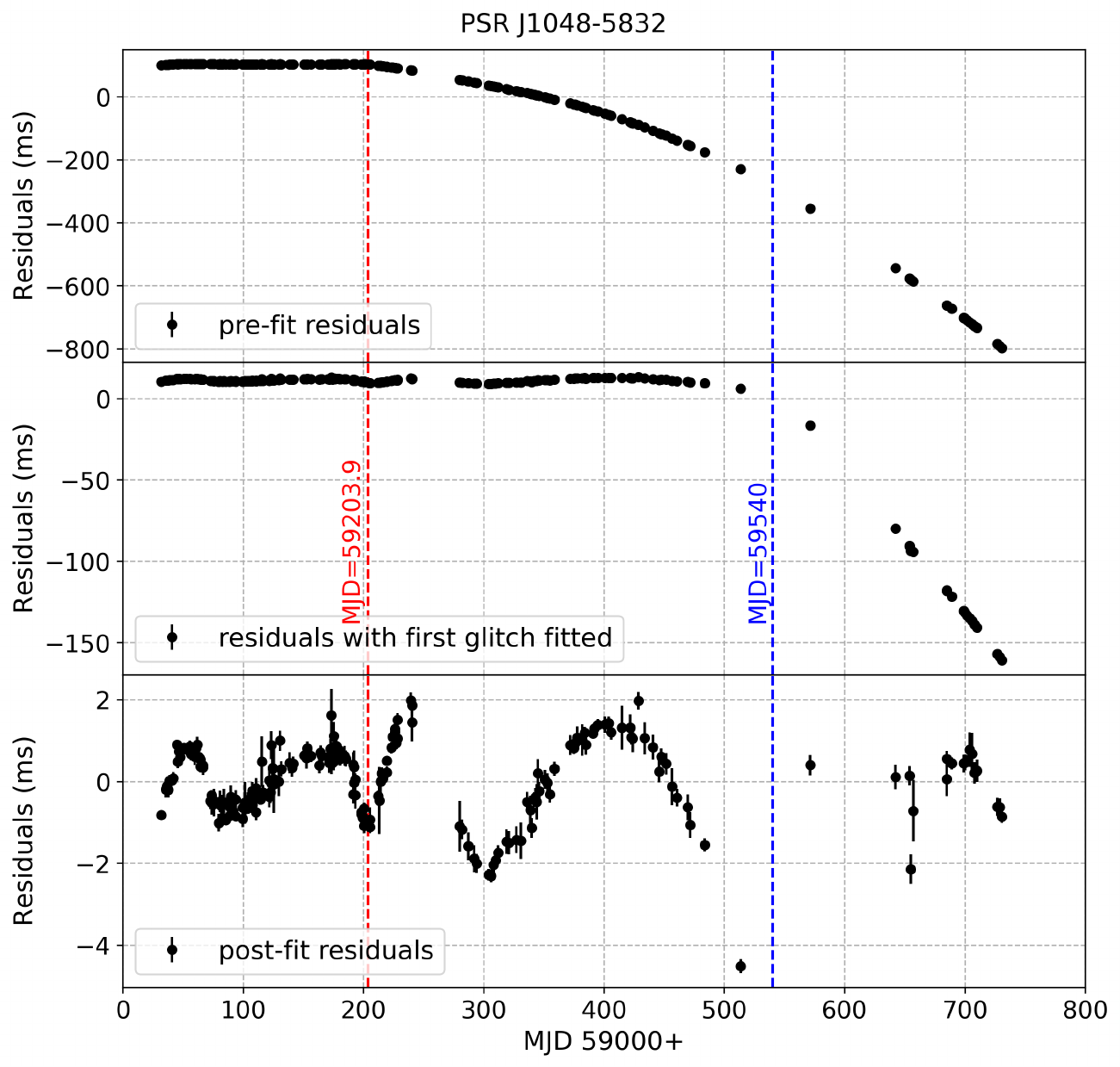}
    \caption{PSR J1048$-$5832 timing residuals for a timing model with no glitches (\textit{top}), with the first glitch included (\textit{middle}) and with the second glitch included (\textit{bottom}). The epochs of the glitches are indicated with coloured vertical lines.}
    \label{fig:1048Residuals}
\end{figure}

In Fig.~\ref{fig:1048Residuals} we show the timing residuals before and after the inclusion of the glitches in the timing model. This timing model is given by Eq.~(\ref{eq:timing-model}) and  Eq.~(\ref{eq:glitch-model}), and the fitted parameters are summarized in Table~\ref{tab:1048parameters}. 
No signs of exponential recovery were found for these glitches, so we do not include the exponential decay term in the final fitting. The white noise in the data was characterised using \texttt{TempoNest} via the parameters $TNGlobalEF$ and $TNGlobalEQ$. We performed a bayesian analysis in a short timespan in order to eliminate the effect of the red noise. We obtained $TNGlobalEF=2.59$ and $TNGLobalEQ=-5.13$; the former indicates the factor by which the template-fitting underestimates the ToA errorbars, and the latter a systematic uncertainty of  $\approx 7~\mu$s.

In order to evaluate the reliability of the reported glitches against timing-noise residuals, we also tested the inclusion of red noise in the timing model. We used the solution obtained by \citet{2020MNRAS.494..228L} for this pulsar\footnote{\url{https://github.com/Molonglo/TimingDataRelease1/}}and re-fitted the timing model without including the putative glitches. We obtained a weighted rms of $Wrms=204~\mu$s and a reduced chi-square of $\chi^2_\mathrm{red}=4.34$. Next, we included the first glitch in the timing model and the residuals decreased significantly, down to \mbox{$Wrms=131~\mu$s} and $\chi^2_\mathrm{red}=1.77$. We then incorporated the second glitch in the model, which led to $Wrms=112~\mu$s and $\chi^2_\mathrm{red}=1.57$. We thus support the interpretation that both events correspond to glitches instead of red noise. 

\begin{table}
  \centering    
  \caption{Parameters of the timing model for PSR J1048$-$5832 and their 1$\sigma$ uncertainties.}
   \begin{tabular}{ccc}
     \hline\hline
     \multirow{2}{*}{Parameter} & \multicolumn{2}{c}{\underline{Value}}  \\
      & glitch 1  & glitch 2  \\[1pt]
     \hline\\[-9pt]
     $\mathrm{PEPOCH}$ (MJD) & \multicolumn{2}{c}{59000} \\
     $\mathrm{F0} (\mathrm{s^{-1}})$ & \multicolumn{2}{c}{8.08166079(4)}\\
     $\mathrm{F1}(\mathrm{s^{-2}})$ &  \multicolumn{2}{c}{$-6.2824(2)\times 10^{-12}$}\\[1pt]\hline\\[-9pt]
     $t_\mathrm{g}$ (MJD)&59203.9(5) & 59540(2)\\
     $\Delta\phi$&$\sim0$& $\sim 0$\\
     $\Delta\nu_\mathrm{p}(s^{-1})$&$7.19(7)\times 10^{-8}$ & $8.02(25)\times 10^{-8}$\\
     $\Delta\dot{\nu}_\mathrm{p}(s^{-2})$& $3.91(9)\times 10^{-15}$ & $1(2)\times 10^{-16}$\\
     \hline
   \end{tabular}
   \label{tab:1048parameters}
 \end{table}
 
In Table~\ref{tab:glitches-anteriores} we recompile the magnitude of all the previous glitches of PSR J1048$-$5832 and compare it with the values of the new glitches reported in this work (on 2020 December 20th and 2021 November 20th). These new glitches can be classified as mini-glitches given that they present values of \mbox{$\Delta \nu_\mathrm{g}/\nu \sim 10^{-8} \ll 10^{-6}$}. We note that there were two small glitches previously detected in this pulsar, but even in these cases their amplitudes were $\approx 3$ times larger than the ones of the two glitches reported in this work.

\begin{table}
    \caption{Magnitude of the glitches in PSR J1048$-$5832. The values for the previous glitches were extracted from the ATNF Catalog \protect\citep{ManchesterATNF2005}.}
    \centering
    \begin{tabular}{c c c}
    \hline \hline\\[-9pt]
        MJD & $\Delta \nu_\mathrm{g} / \nu$ $(10^{-9})$ & References \\
    \hline
48944(2)&	25(2)& \cite{Wang:2000nw}\\
49034(9)&	2995(7)& \cite{Wang:2000nw}\\
50788(3)&	771(2)& \cite{Wang:2000nw}\\
52733(37)&	1838.4(5)& \cite{Yu2013}\\
53673.0(8)&	28.5(4)& \cite{Yu2013}\\
54495(4)&	3044.1(9)& \cite{Lower:2021rdo}\\
56756(4)&	2964(3)& \cite{Lower:2021rdo}\\
59203.9(5)&  8.89(9) & This work\\
59540(2)&  9.9(3) & This work\\
    \hline
    \end{tabular}
    \label{tab:glitches-anteriores}
\end{table}


\subsection{Glitch detection in PSR J0835$-$4510 (Vela)}\label{sec:J0835}

We first reported the detection of a new $(\#22)$ glitch in Vela in \cite{2021ATel14806....1S} (the 21 glitches previously reported are listed in the
ATNF catalogue {\url{http://www.atnf.csiro.au/people/pulsar/psrcat/glitchTbl.html}}).
We observed the Vela pulsar on Jul 21 for $165$ min with A1 and $206$ min with A2 (MJD 59416.6321--59416.7666). We measured a barycentric period of $P_\mathrm{bary} = 89.4086241(17)$~ms, consistent with the pulsar ephemeris at that time. No glitch was observed during that observation. In our following observation on Jul 22 (started in MJD 59417.6549) with A2, we obtained a period $P_\mathrm{bary} = 89.4065093(15)$~ms, showing a decrease of $\Delta P = 0.113~\mu$s with respect to the expected period, which corresponds to \mbox{$\Delta P/P = 1.26\times10^{-6}$}. This result was confirmed with a subsequent observation on Jul 23 with A1 and A2. This first analysis placed the new Vela glitch between MJD 59416.7666--59417.6549. Subsequent reports \citep{2021ATel14807....1D,2021ATel14808....1O,2021ATel14812....1S} narrowed the glitch epoch to MJD 59417.618--59417.628.

\begin{figure}
    \centering
    \includegraphics[width=\linewidth]{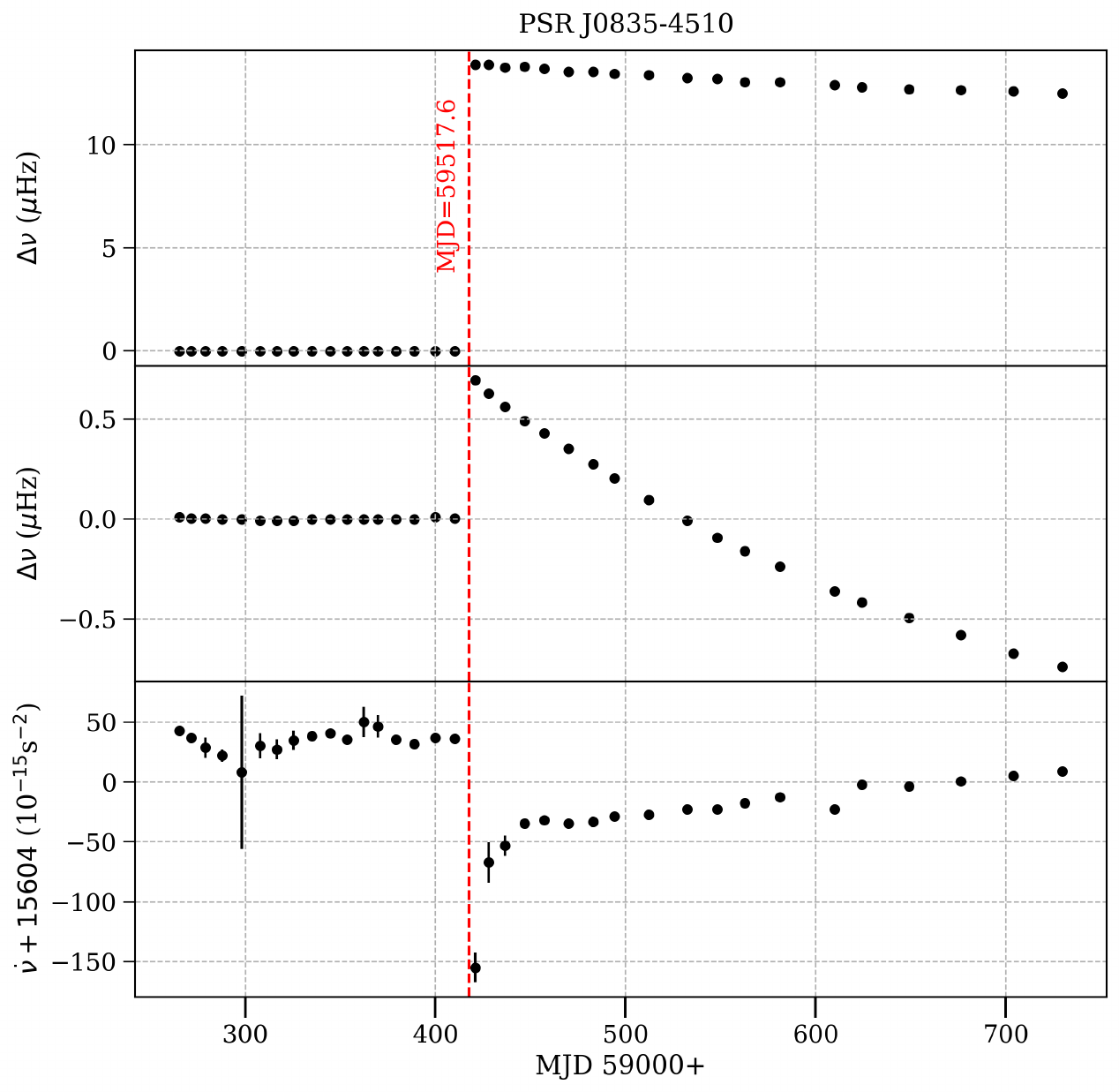}
    \caption{Timing analysis of Vela's glitch. \textit{Top}: variations in the rotational frequency $\Delta\nu$ relative to the pre-glitch solution. \textit{Center}: an expanded plot of $\Delta\nu$, in which the mean post-glitch value has been subtracted from the post-glitch data. \textit{Bottom}: variations of the frequency first derivative $\Delta\dot{\nu}$. The vertical dashed line marks the glitch epoch.}
    \label{fig:VelaPlugin}
\end{figure}

Here we present a more thorough analysis of the Vela timing behaviour around the epoch of the glitch. In Fig.~\ref{fig:VelaResiduals}a) we show the residuals before including the glitch in the timing model. We focused on a time window of roughly 90 days centered in the glitch epoch (MJD 59417.6). During the pre-glitch window (MJD 59371.7--MJD 59416.7) our restricted dataset includes observations in 21 days with A1 and in 27 days with A2, while during the post-glitch window (MJD 59418.7–-MJD 59463.6) we have observations in 30 days with A1 and in 23 days with A2. 

\begin{figure}
    \centering
    \includegraphics[width=\linewidth]{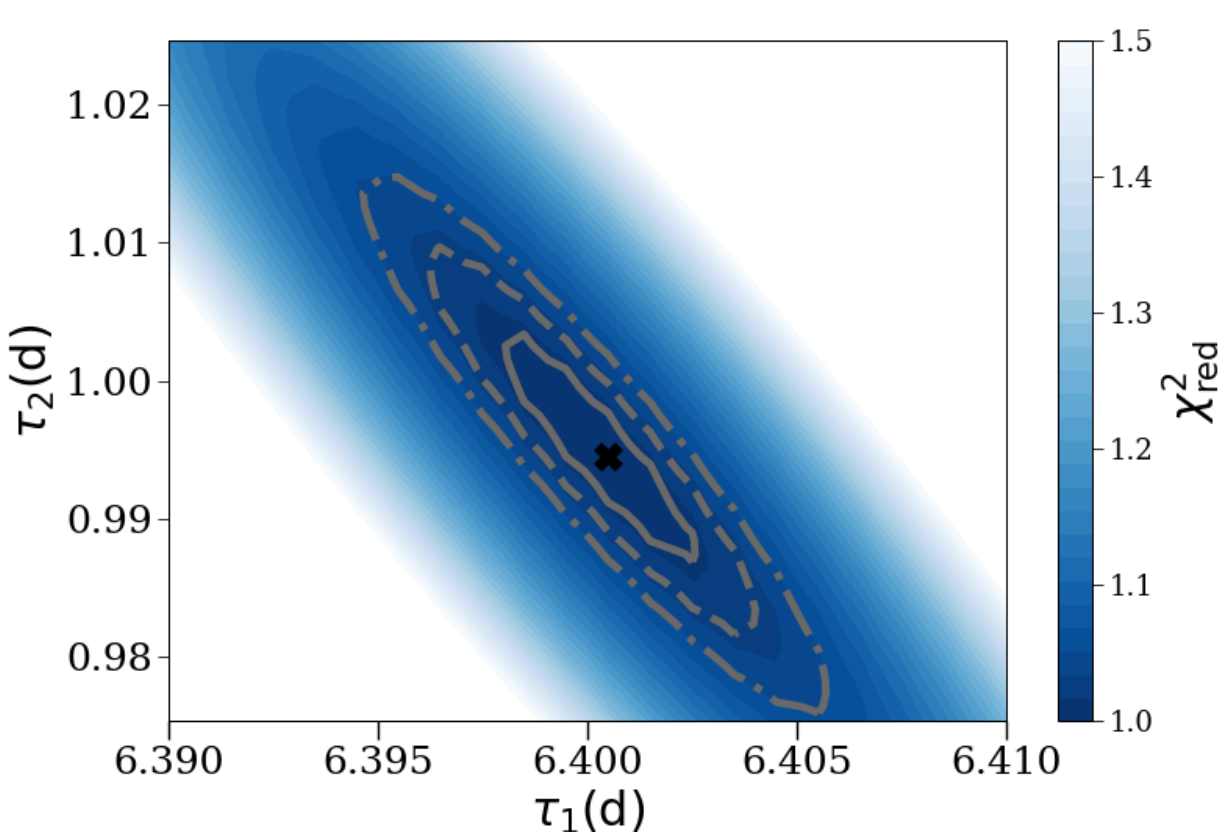}
    \caption{Best fit of the decay time constants $\tau_1$ and $\tau_2$ for the 2021 Vela glitch. The solid line, dashed line, and dot-dashed line indicate the 1-, 2- and 3-$\sigma$ confidence regions.}
    \label{fig:VelaTau}
\end{figure}

We first derived the rotational parameters of the timing model before and after the glitch by fitting $\nu$, $\dot\nu$ and $\ddot\nu$ in Eq.~(\ref{eq:timing-model}) to the pre-glitch and post-glitch data. For this we excluded the ToAs within 10 days after the glitch in order to avoid the effects of the strong exponential decay shown in Fig.~\ref{fig:VelaPlugin}. By comparing the results for the pre-glitch solution and post-glitch asymptotic solution, we estimated the parameters $\Delta\nu_\mathrm{p}$, $\Delta \dot{\nu}_\mathrm{p}$ and $\Delta \Ddot{\nu}$. The residuals after including and fitting these parameters in the timing model are shown in Fig.~\ref{fig:VelaResiduals}b). 

The high cadence of observations of this pulsar makes it possible to monitor the recovery process rigorously. We used the glitch plugin in \texttt{TEMPO2} to obtain values of $\nu$ and $\dot{\nu}$ from individual sections of data, with each section spanning $\sim10$~d (Fig.~\ref{fig:VelaPlugin}). Both the glitch plugin and the timing residuals in Fig.~\ref{fig:VelaResiduals}b) clearly indicated a decaying term of a few days. We then searched for the value of the decay timescale $\tau_{d1}$ that minimised the reduced chi square of the timing residuals, $\chi^2_\mathrm{red}=\chi^2/dof$, with $dof$ the number of degrees of freedom of the model. For this we explored systematically different values of $\tau_{d1}$, starting with a scarce sampling over a broad range of values between $0$~d and $100$~d with a $1$~d step, obtaining $\tau_{d1}\sim6$~d. We then progressively iterated on smaller ranges and smaller steps. For the final run we used a step of 0.001~d over the range of 6.3--6.5~d. For each fixed value of $\tau_{d1}$, we fitted $\Delta \nu$, $\Delta \dot\nu$, $\Delta \Ddot\nu$ and $\Delta \nu_\mathrm{d}$, and obtained the corresponding $\chi^2_\mathrm{red}$. With this procedure we obtained $\tau_\mathrm{d1}= 6.39(1)$~d. The residuals, shown in Fig.~\ref{fig:VelaResiduals}c), suggest the existence of an additional decay term. We therefore explored systematically the values of both decay timescales as explained before. The results are shown in Fig.~\ref{fig:VelaTau}. The error bars at $1\sigma$, $2\sigma$ and $3\sigma$ were calculated as the $\tau_\mathrm{d1}$ values that increase the $\chi^2_\mathrm{red}$ by $\Delta \chi^2 /dof = K$, with $K=2.30$, $6.17$ and $11.8$,  respectively \citep{Numerical_recipes}. The fitted glitch parameters are given in Table~\ref{tab:Vglitch}. For this analysis the white noise was characterised using \texttt{TempoNest} similarly as it was done with J1048$-$3832 (Sect.~\ref{sec:J1048}), obtaining $TNGlobalEF=3.95$ and $TNGLobalEQ=-5.3$. Finally, in Fig.~\ref{fig:VelaResiduals}d) we show the post-fit residuals after including all the parameters in the timing model given by Eq.~(\protect\ref{eq:glitch-model}). 

The glitch epoch $t_\mathrm{g}$ is consistent with the reports mentioned before. It can be seen $t_\mathrm{g}$ is accurate because $\phi_\mathrm{g}\sim 0$. $Q_1=0.2(1)\%$ and $Q_2=0.7(1)\%$ indicates that the glitch process is dominated by the permanent jump in the frequency, as commonly detected in large glitches. 
We have used the values of $Q_1=0.2(1)\%$ and $Q_2=0.7(1)\%$ (fraction of glitch recovery), $\tau_1=6.400(2)$ and $\tau_2=0.994(8)$ (decay time) for this 2021 Vela glitch to compare to all other available glitches in ATNF catalog with one, two or four decay rates as displayed in Fig.~\ref{fig:Qtau}. 

\begin{figure}
    \centering
    \includegraphics[width=\linewidth]{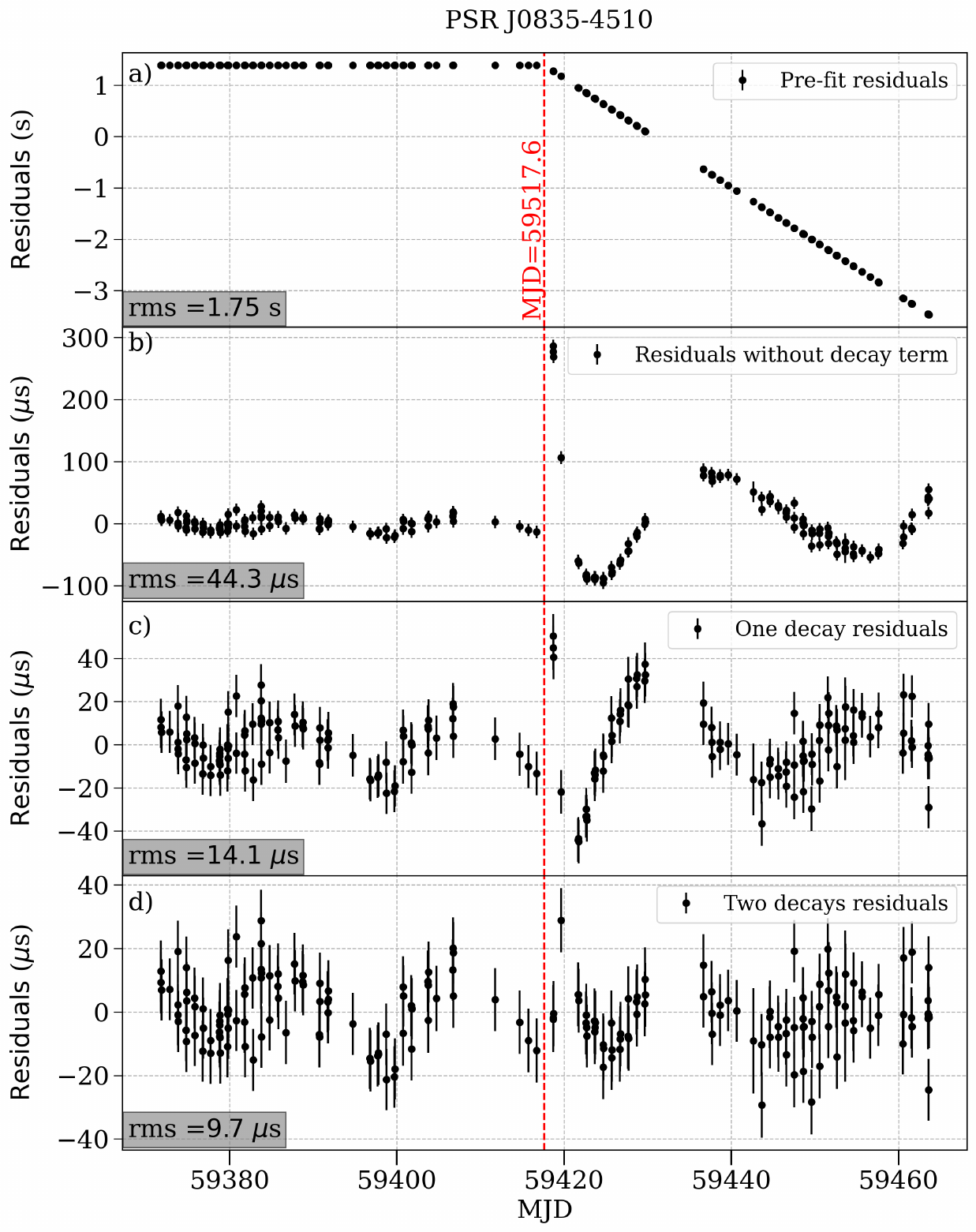}
    \caption{Vela's timing model with the parameters from Table~\ref{tab:Vglitch}.} 
    \label{fig:VelaResiduals}
\end{figure}

\begin{table}
  \centering    
  \caption{Parameters of the timing model for the 2021 July 22nd Vela glitch and their 1$\sigma$ uncertainties.}
   \begin{tabular}{ll}        
     \hline
     Parameter & Value \\
     \hline
     $\mathrm{PEPOCH}$ (MJD)&59417.6193 \\
     $\mathrm{F0}(\mathrm{s^{-1}})$& 11.18420841(1)\\
     $\mathrm{F1}(\mathrm{s^{-2}})$ &  $-1.55645(4)\times 10^{-11}$\\
     $\mathrm{F2}(\mathrm{s^{-3}})$ &  $6.48(1)\times 10^{-22}$\\
     $\mathrm{DM}(\mathrm{cm^{-3} pc})$ & 67.93(1)\\
     $t_\mathrm{g}$ (MJD) &59417.6194(2)\\
     $\Delta\nu_\mathrm{p}$ (s$^{-1}$) &1.381518(1) $\times 10^{-5}$\\
     $\Delta\dot{\nu}_\mathrm{p}$ (s$^{-2}$)&$-8.59(4)\times 10^{-14}$\\
     $\Delta\ddot{\nu}$\quad (s$^{-3}$)&$1.16(3)\times 10^{-21}$\\
     $\Delta\nu_\mathrm{d1}$\, (s$^{-1}$)&$3.15(12)\times 10^{-8}$\\
     $\tau_\mathrm{d1}$ (days) & 6.400(2)\\
     $\Delta\nu_\mathrm{d2}$\, (s$^{-1}$)&$9.9(6)\times 10^{-8}$\\
     $\tau_\mathrm{d2}$ (days) & 0.994(8)\\
     $\Delta \phi$ & $\sim0$\\
     $\Delta\nu_\mathrm{g}/\nu$ & $1.2469(5)\times 10^{-6}$\\
     $\Delta\dot\nu_\mathrm{g}/\dot\nu$ & 0.084(5)\\
     $Q_1$ & 0.00226(9)\\
     $Q_2$ & 0.0071(4)\\
     \hline
   \end{tabular}
  \label{tab:Vglitch}
 \end{table}

\begin{figure}
    \centering
    \includegraphics[width=\linewidth]{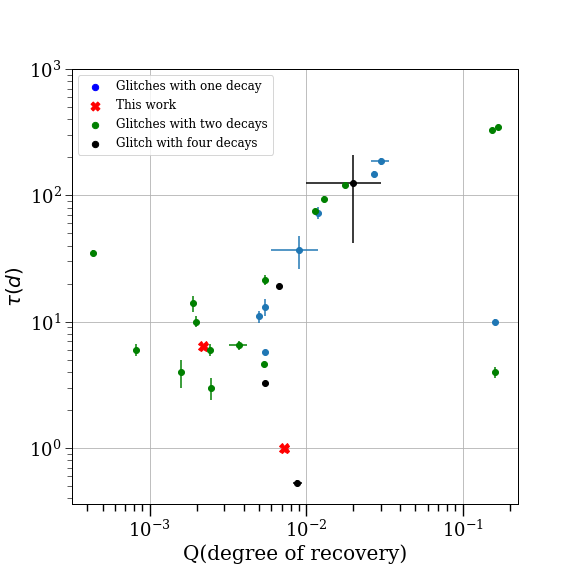}
    \caption{Comparison of current and previous glitches decaying parameters for Vela pulsar.}
    \label{fig:Qtau}
\end{figure}

\subsection{Glitch validation in PSR J0742$-$2822}\label{sec:J0742}

PSR J0742$-$2822 (PSR B0740-28) had a total of eight glitch events reported
\footnote{\url{http://www.jb.man.ac.uk/pulsar/glitches/gTable.html}, 
\url{https://www.atnf.csiro.au/people/pulsar/psrcat/glitchTbl.html}.},
with the latest \#8 found on MJD 56725.2(2) \citep{2022MNRAS.510.4049B}. 
The largest glitch reported by \citep{2011MNRAS.414.1679E}
was \#7 with a $\Delta \nu / \nu = 92(2) \times 10^{-9}$
and $\Delta \dot\nu / \dot\nu = -0.372(96)$


On 2022 September 21, MJD=59839.4(5), a new glitch \#9 in PSR J0742$-$2822 was reported by
\citep{2022ATel15622....1S}. We have been able to confirm this glitch with our data \citep{2022ATel15638....1Z} and find relative jumps of $\Delta \nu / \nu = 4.29497(2) \times 10^{-6}$ and $\Delta \dot\nu / \dot\nu = 0.0510(7)$, making it the largest recorded glitch for this pulsar,
but due to the sparcity of our data around the glitch date we are unable to search for any putative exponential decay component. 

\subsection{A new glitch detection in PSR J1740$-$3015}\label{sec:J1740}

PSR J1740$-$3015 (PSR B1737$-$30) is one of the most frequently glitching pulsars known, 
with 37 recorded in \url{https://www.atnf.csiro.au/people/pulsar/psrcat/glitchTbl.html}, 
with a large variety of jump amplitudes, ranging from 
$\Delta \nu / \nu $ as small as $10^{-9}$ to as large as
$2.66\times 10^{-6}$  \citep{2022MNRAS.510.4049B}. 

On 2022 December 22, MJD=59935.1(4), we detected a new glitch in PSR J1740$-$3015 that was reported in 
\cite{2023ATel15838....1Z} and confirmed by UTMOST \citep{2023ATel15839....1D} and uGMRT \citep{2023ATel15851....1G}.
We found a relative jump of $\Delta \nu / \nu = 3.32(3) \times 10^{-7}$ and plan 
to continue monitoring PSR J1740$-$3015 to improve the post-glitch timing solution.


\section{Analysis Methods: Pulse-by-pulse analysis of the 2021 Vela glitch}\label{sec:analysis}

In this section we report the analysis of the observations around the Vela glitch pulse by pulse. 
High-resolution single-pulse micro-structure pulse studies of the Vela pulsar were reported in \cite{Kramer:2002us}, while the temporal evolution of the pulses for large time-scales was studied in \cite{Palfreyman+2016}.
Here we take advantage of the large amount of our daily data well suited for statistical and machine learning studies. 
Our approach has been carried out using a combination of the VAE reconstruction and the SOM clustering techniques. 

We analyze five observations on 2021, July, 19th, 20th, 21st, 23rd and 24th, performed all with antenna A1 configuration on a single polarization at 112~MHz bandwidth. 
The number of pulses in each observation is given in Table~\ref{tab:observations}. Those are uninterrupted single observations with A1 and
we supplement them with antenna A2 observation for July 20 and July 22, the day of the glitch, which are split into two and three
observations respectively, as show in Table~\ref{tab:A2}.
All observations considered here are folded with a fixed  $DM=67.93(1)$~pc\,cm$^{-3}$  from the ATNF catalogue\footnote{\url{http://www.atnf.csiro.au/people/pulsar/psrcat/}} (as we have seen very small variations during each observation, $DM<0.2$~pc\,cm$^{-3}$)
and cleaned from radio frequency interferences using the code \texttt{RFIClean} \citep{Maan2020} with protection of the fundamental frequency of Vela (11.184~Hz). The complete procedure is described in Appendix C of \cite{Lousto:2021dia},
where we found that using \texttt{rfifind} \citep[a task within \texttt{PRESTO};][]{PRESTO} on the data output from \texttt{RFIClean} further improves the S/N in most of the cases we studied. The amplitudes of the pulses are in arbitrary units as we did not observe any flux calibrator. Their relative distribution, day per day analyzed here (Table~\ref{tab:observations}), is displayed in Fig.~\ref{fig:17New}. This figure also includes the added two A2 observations of the glitch day 2021 July 22 (Table~\ref{tab:A2}). We note the qualitative similarities of the A1 pulse distributions pre-glitch on top, while the post-glitch observations are a bit more heterogeneous. For a more quantitative comparison one can look at the clusters parameters in the Tables in Appendix \ref{sec:appendix}.

\begin{table*}
	\centering
	\caption{Date of each observation with A1, duration in hours, the MJD at the beginning and end of the observations, the corresponding number of single pulses, instantaneous topocentric period, $P_{\mathrm{obs}}$, and estimated signal to noise ratio (SNR) for the selected observations around the 2021 Vela glitch used for the pulse-by-pulse analysis. The estimated time of the glitch on July 22 is MJD 59417.6194(2).}
	\label{tab:observations}
	\begin{tabular}{l c l l r l r} 
		\hline
		Date & Duration [h] & initial MJD & final MJD & \# pulses & $P_{\mathrm{obs}}$ [ms] & SNR\\
		\hline
		July 19 & 3.55 & 59414.6256  & 59414.7737 & 143082 & 89.4142714 & 265.9\\
		July 20 & 2.45 & 59415.6688  & 59415.7708 & 98545 & 89.4142431 & 241.0\\
		July 21 & 2.45 & 59416.6656  & 59416.7680 & 98948 & 89.4141939 & 372.3\\
\hline
		July 23 & 2.20 & 59418.6708 & 59418.7626 & 88740 & 89.4139894 & 283.3\\
		July 24 & 0.33 & 59419.6238 & 59419.6377 & 13401 & 89.4139192 & 69.0\\
		\hline
	\end{tabular}
\end{table*}

\begin{figure*}
   \centering 
   \includegraphics[width=\textwidth]{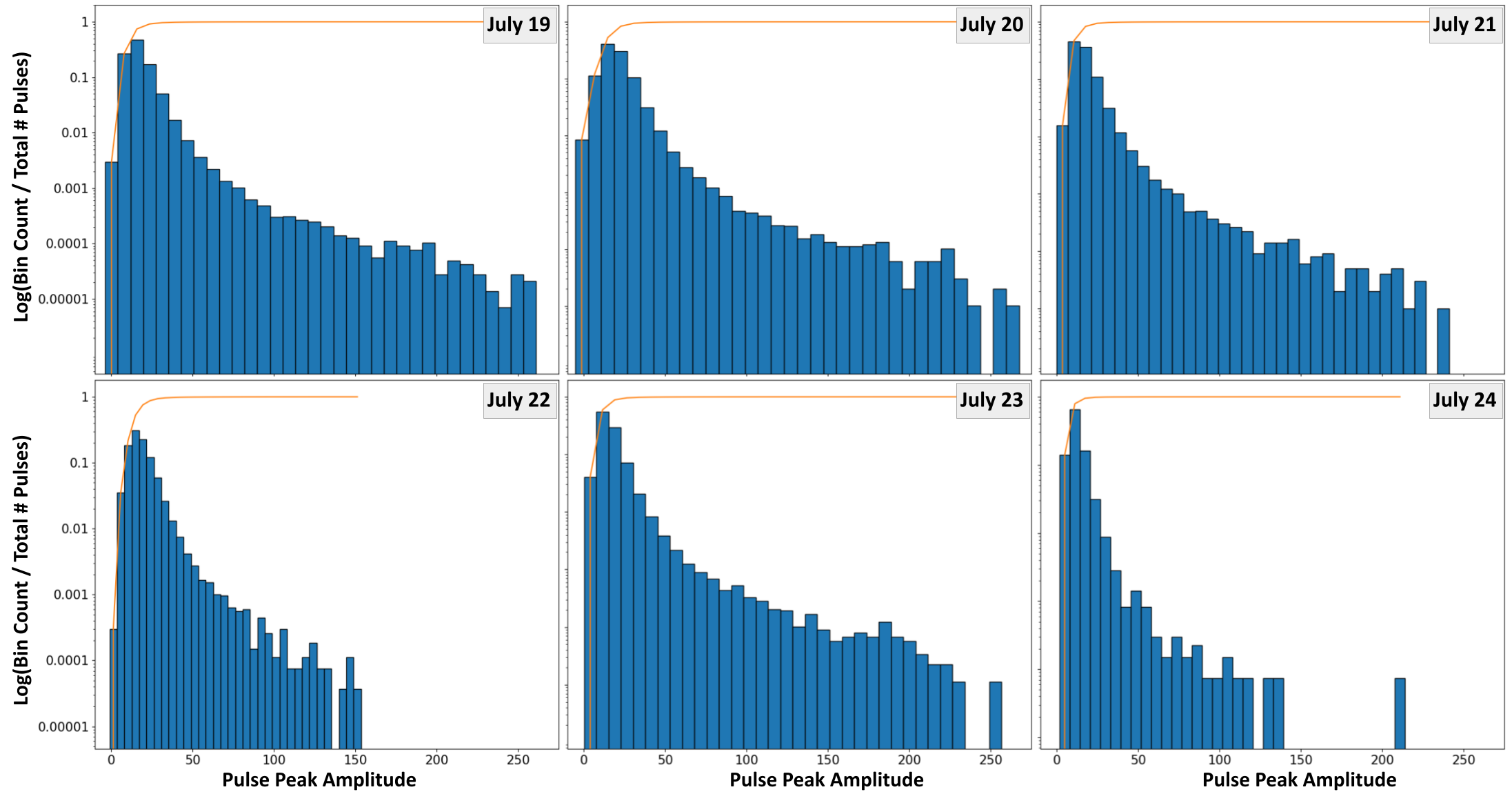}
     \caption{Peak amplitude of single pulses distribution for observations with A1 on 2021, July 19, 20, 21, 23, and 24, and July 22 with A2. The top curve is the cumulative sum.}
  \label{fig:17New}
 \end{figure*}

\subsection{Self-Organizing Map (SOM) techniques}\label{sec:CS}
Here we describe a deep learning generative and clustering method built on Variational AutoEncoders (VAE) and Self-Organizing Maps (SOM) to perform Vela per-pulse clustering in an unsupervised manner. Recently, deep learning has been leveraged across many domains -- from medical imaging tasks to natural language translation -- with related astronomical tasks of galaxy image denoising \citep{Chianese:2019ifk}. With deep neural networks, latent representations can be learned via the hierarchical information bottlenecking intermediate layers that capture the inherent feature characteristics of the input data. From these latent representations, one can efficiently group the individual pulses into hierarchically meaningful clusters. Clusters described here within refer to the automatic grouping of similar signals based on the learned underlying latent structure of the data and a defined distance measure. It requires no derived physical parameters or prior knowledge of relationships between data points. Specifically, the VAE takes in the raw pulsar signal and the SOM takes in either the VAE's latent representation $\mathbf{z}$ or its reconstructed data signal $\mathbf{\hat{x}}$.

For the task of de-noising the pulsar signals and generating a meaningful latent representation, we resort to the popular unsupervised approach of the Variational Autoencoder, a deep learning framework that reconstructs a given input after being subjected to dimensionality regularization and stochasticity \citep{kingma2014autoencoding}. We refer to \citep{Lousto:2021dia} for mathematical details and present a methodological overview instead. For each pulse, $\mathbf{x_i}$, its mean, $\mu$, and standard deviation, $\sigma$, are generated from a neural network encoder and a latent sample $\mathbf{z_i}$ is derived from its variational approximation $q_{\phi}(\mathbf{z_i}|\mathbf{x_i})$ of a Gaussian distribution. This is then passed through an identical but reversed neural network decoder to get the reconstructed output $\mathbf{\hat{x}_i}$ and the error in reconstruction is leveraged as an optimization objective. The information bottleneck allows the network to capture only the meaningful variations within the data distribution, encoded within the dimensions of the latent space, and discard any irrelevant noise. The stochastic nature of the \textit{variational} approach encourages the encoding network to learn a structurally meaningful latent distribution, such that 'walks' in the latent space produce interpretable interpolations between data points or across features.

Once the de-noising VAE is trained, we perform pulse clustering through the Self-Organizing Map (SOM), a neural network-based clustering algorithm that optimises a two-dimensional discrete map to topographically represent the input data as nodes \citep{teuvo1988som}. It is, in essence, a generalised form of the K-Means algorithm, in which the 'centroids' exert topographical force on its neighbours whenever it is updated. The SOM consists of a 2D grid containing $M$ nodes, $\mathcal{V}=\{v_1,v_2,...,v_{M}\}$, that, for each node $v\in\mathcal{V}$, have assigned weight vectors $\mathbf{r}^{v}$. The grid is iteratively optimised to minimise the Euclidean distance between every input and its closest node called the Best Matching Unit (BMU) by dragging the node towards the input. To preserve the SOM's topographic structure, updated nodes pull its neighbouring nodes in its update direction - often done with a neighbourhood distance weight function that decays over the course of fitting. Training completes when the relative change in error between iterations stalls and the resulting node positions represent cluster centres (or \textit{prototypes}) of the input and new samples can be assigned to the closest prototype. Though both the latent representations or the original, noisy signals can be used as inputs to the SOM, we primarily consider the reconstructed signals $\mathbf{\hat{X}}$ as they are sufficient approximations to the original and minimise noise (samples are provided in Fig.~\ref{fig:7-20pulses}). 

To recap the method simply, we employ a two-stage process where the raw noisy pulses are first de-noised (VAE) and then are grouped into clusters second (SOM). The raw noisy pulses $\mathbf{X}$ are denoised into smooth approximations $\mathbf{\hat{X}}$ through neural networks that compress the input into a lower-dimensional stochastic space and then try to reconstruct the signal. We then define a 2D grid of $M$ nodes, $\mathbf{V}_{1:M}$, each initialised as a random vector in data space. The grid is iteratively updated through a competitive process where the input signals are presented to all nodes and the closest node via a distance measure (e.g. Euclidean distance) is chosen as the 'best matching unit'. This node and its grid neighbours are then slightly pulled closer to that input data point. This process is repeated until the grid is stable. The result is a set of cluster centres and assignments that partition similar signals into groups based on the dataset's latent structure.
The schematic diagram of VAE and usage of SOM for clustering is presented in Fig.~11 of \cite{Lousto:2021dia}.

\subsection{Results}\label{sec:results}

We have collected the results of the SOM clustering for the five days of observation in Fig.~\ref{fig:All}.
The results are displayed by days in successive rows and the three columns correspond to the choice of collecting the whole set of pulses in 4, 6, and 9 clusters respectively. The glitch on 2021 July 22nd would lie between rows 3 and 4. 
We have chosen the same vertical scale to represent the mean pulse of each cluster over the choices of the number of clusters and over the days of observation in order to exhibit the relative amplitudes, also affected by the different amount of observing time.
Figures \ref{fig:All}, \ref{fig:7-22A2}, \ref{fig:7-20}, \ref{fig:7-20pulses}, 
and \ref{fig:7-20A2} display pulses amplitudes (in the arbitrary units
coming from the \texttt{PRESTO} (FFT) normalization). We have not used standard sources to seek a normalization of the observations, although we provide 
the signal-to-noise ratio (SNR) of each observation as provided by \texttt{PRESTO} in Table~\ref{tab:observations}.

The labelling of the clusters in each panel are ordered from the largest to the lowest amplitude mean pulse; while cluster 0 is the total mean pulse of the whole observation and remains the same over the three horizontal panels as a reference value.
We first note an increase in the amplitude of the mean pulse of the cluster 1 as we increase the numbers of clusters allowed to SOM. They also decrease the number of pulses per cluster (as expected), what explains the increase in amplitude. This behaviour is shared by clusters 2 and 3 and successively. We also note an earlier arrival and a mild decrease in the width of the high amplitude clusters (feature that could be used for improved timing in other circumstances or for other millisecond pulsars as we noted in \cite{Lousto:2021dia}).
These points are more precisely quantified, with estimated errors, in the Tables in Appendix~\ref{sec:appendix}.

We note that the cluster distribution follows a similar pattern to our previous studies with observations about six months before this glitch, on 2021 January 21, 24, 28 and March 29. However, the observations of July 20, two days before the glitch, show a baseline behaviour with unusual activity before and after the main pulse, which in turn decreases in amplitude relative to other neighbouring days. On the other hand this effect gets suppressed when the observation is not dissected into clusters.

\begin{figure*}
   \centering 
   \includegraphics[width=\textwidth]{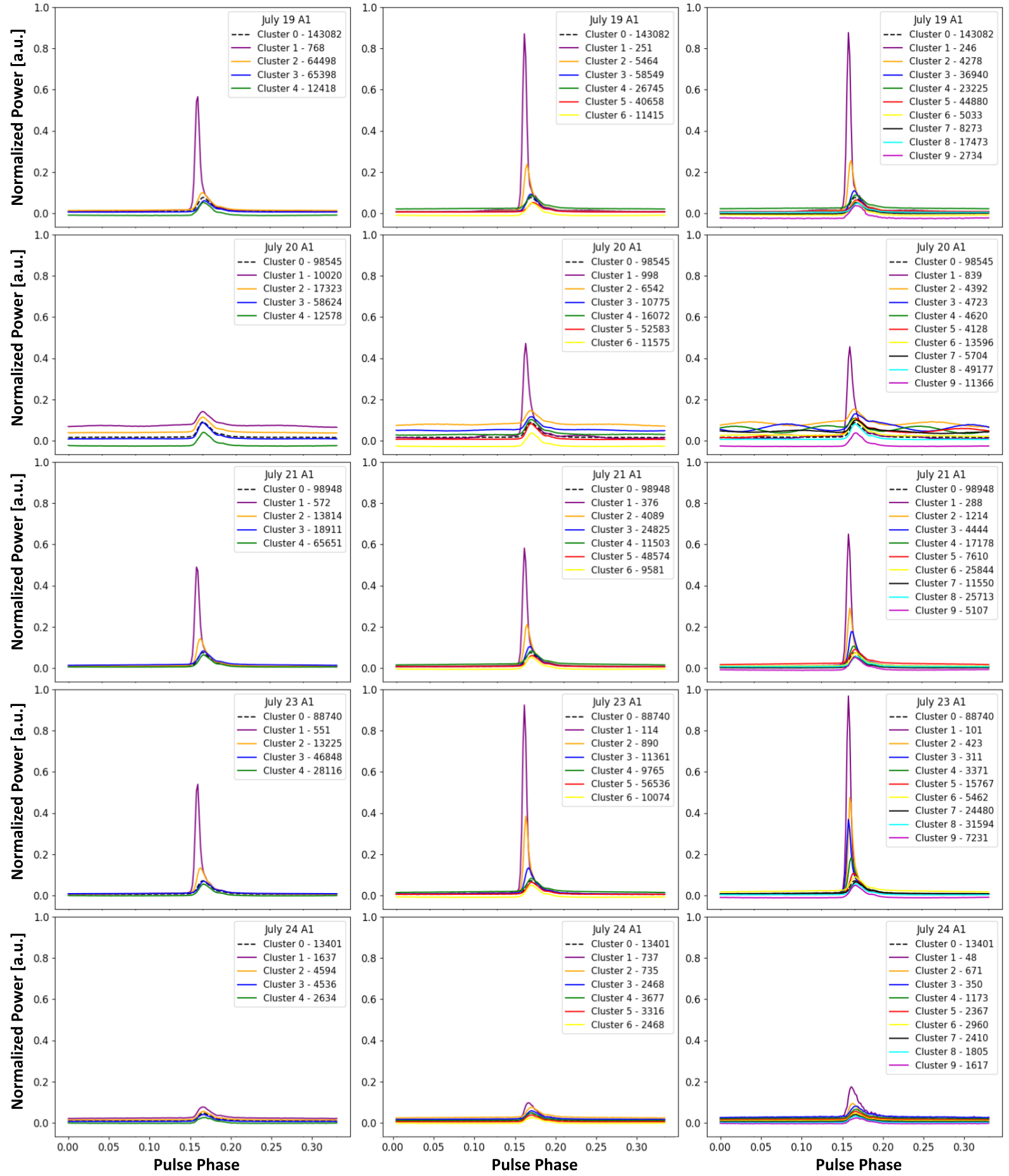}
     \caption{
     Mean cluster reconstruction for observations with A1 on 2021, July 19, 20, 21, 23, and 24, using 4, 6, and 9 SOM clustering. 200 (out of total 611) phase bins were taken around the mean peak of each day to perform the single-pulse analysis on.}
  \label{fig:All}
 \end{figure*}

In Figs.~\ref{fig:Time4}-Figs.~\ref{fig:Time9} we represent the sequence of pulses for each day of observation with large amount of data: 19, 20, 21, and 23 July (rows) per SOM clustering for 4, 6, and 9 clusters (columns) in blocks of ordered 5000 pulses, labelled by an integer number index. Those histograms provide a rough distribution over time of the clusters during each observation. The 4 clusters distribution gives a more robust view of the classes of pulses with a certain consistency over time except for the second half of the 20 July observation where there seems to be a shuffle of the high amplitude pulses into the low amplitude ones or an increase of the general noise of the signal.
The 6 and 9 SOM clusters decomposition confirms in more detail these findings. During the July 20 observation there is a transition from a high amplitude to a low amplitude dominated number of pulses that is then recovered in the posterior days of observation. 

\begin{figure*}
     \includegraphics[width=0.99\textwidth]{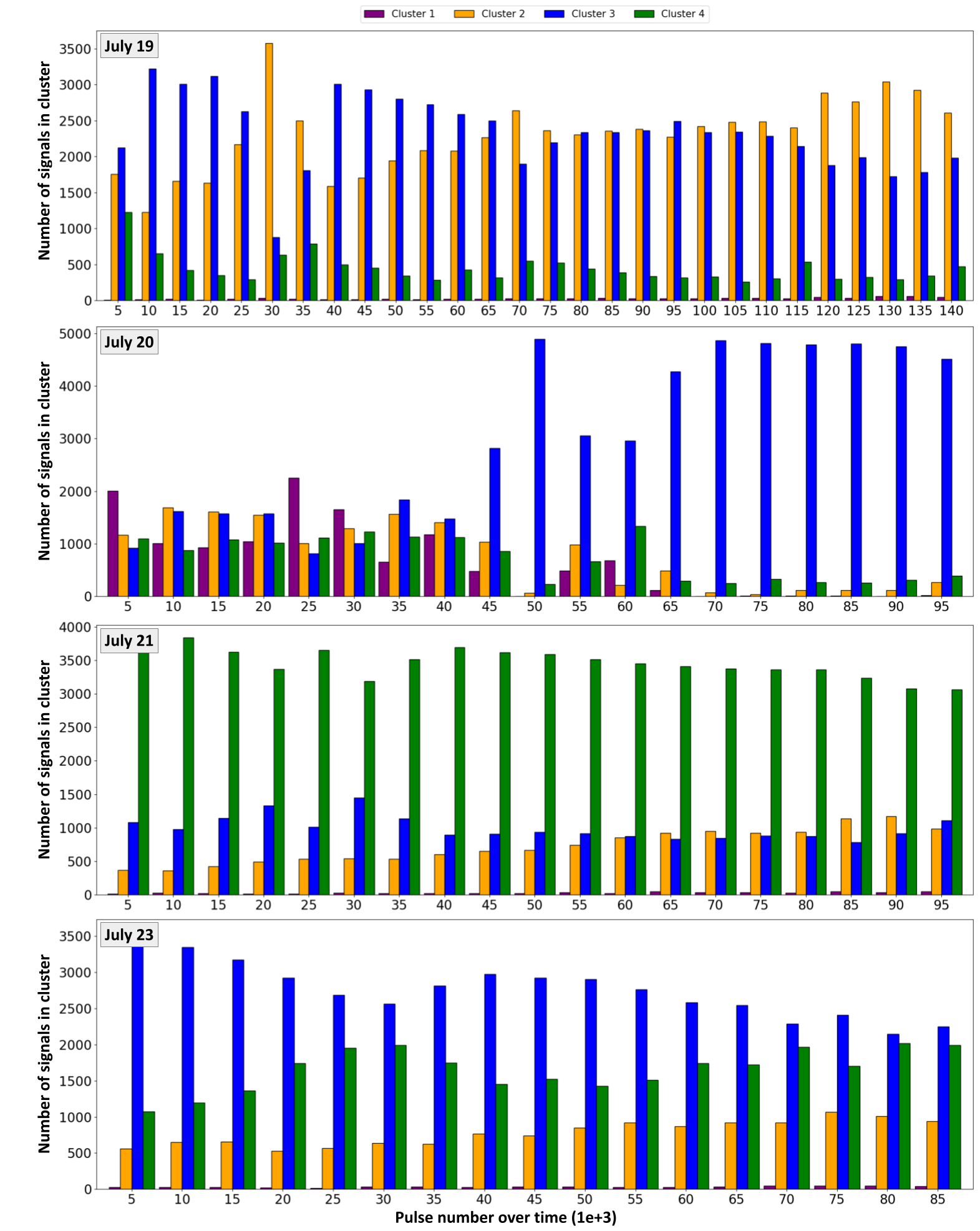}
     \caption{Time distribution of clusters (binned every 5000 pulses) for 2021, July 19,20,21,23 observations on Antenna 1 for 4 SOM clustering.} 
  \label{fig:Time4}
\end{figure*}


\begin{figure*}
     \includegraphics[width=0.99\textwidth]{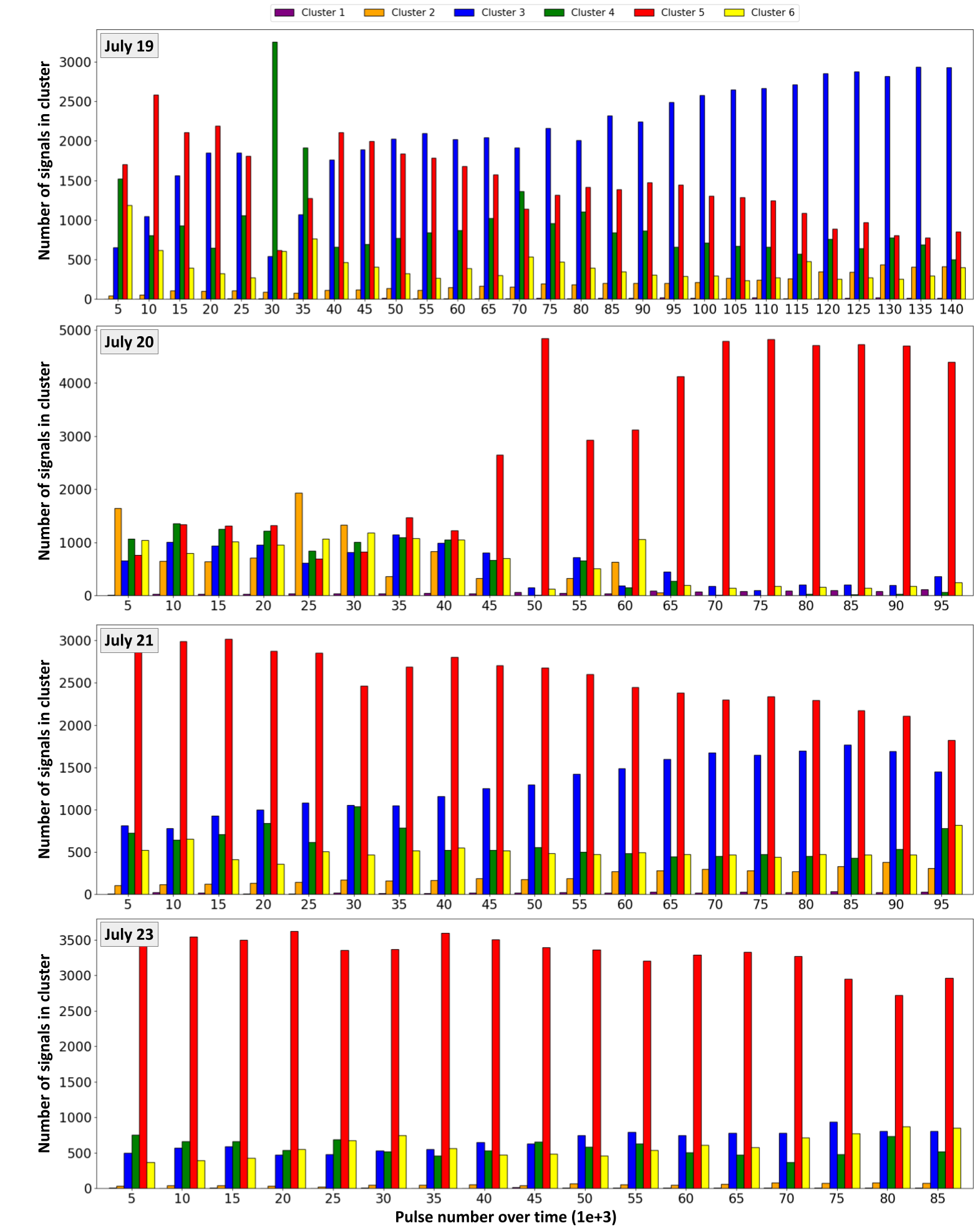}
     \caption{Time distribution of clusters (binned every 5000 pulses) for 2021, July 19,20,21,23 observations on Antenna 1 for 6 SOM clustering.} 
  \label{fig:Time6}
\end{figure*}


\begin{figure*}
     \includegraphics[width=0.99\textwidth]{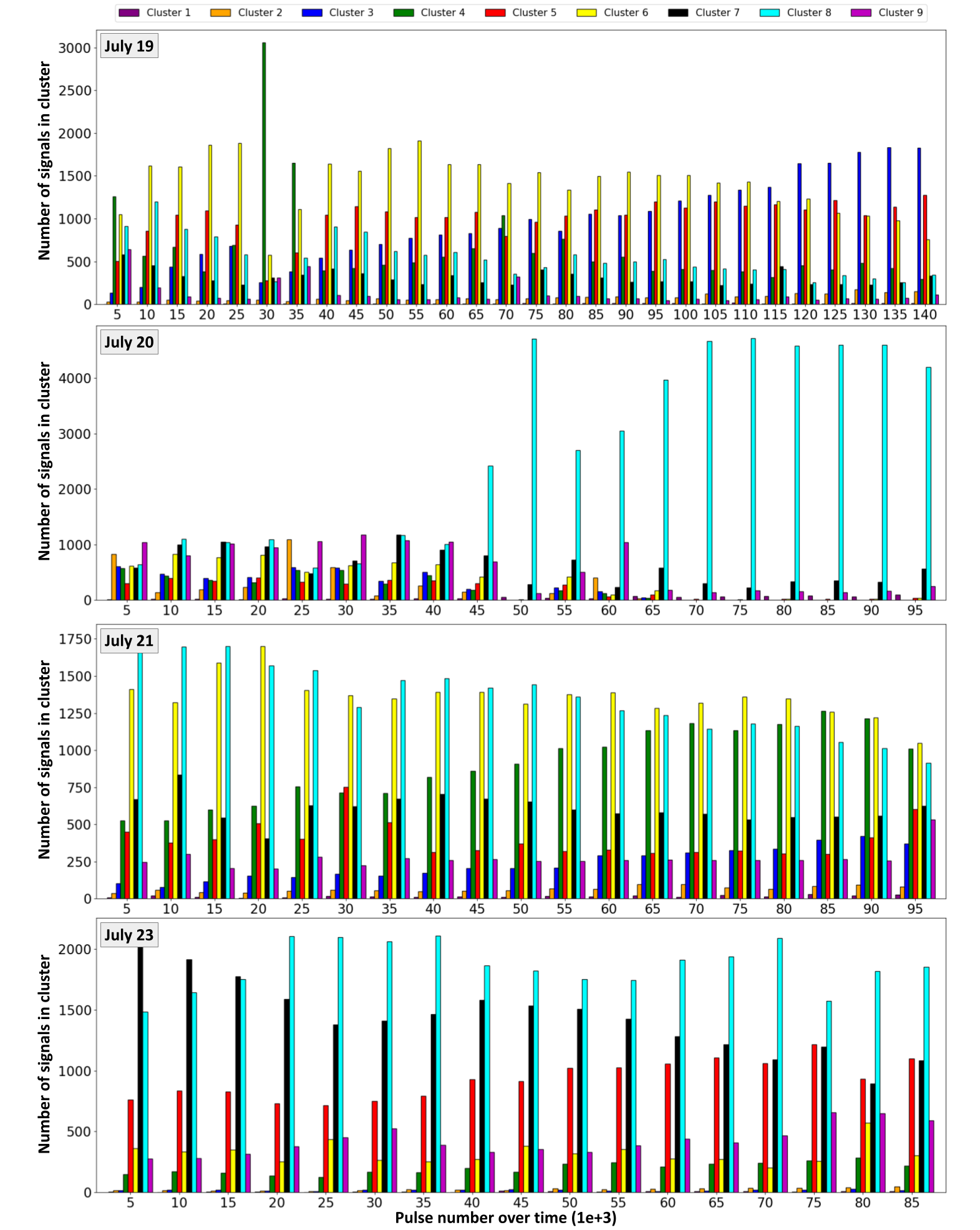}
     \caption{Time distribution of clusters (binned every 5000 pulses) for 2021, July 19,20,21,23 observations on Antenna 1 for 9 SOM clustering.} 
  \label{fig:Time9}
\end{figure*}




\subsubsection{Glitch Day:  2021, July 22 observations with A2}\label{sec:July22}

Unlike the continuous observations with A1, those performed with A2 suffered from short (a few seconds) interruptions due to some software/hardware limitations. The observations on  2021 July 22 (the day of the glitch) are divided in three parts as described in Table~\ref{tab:A2}. The first of those observations, starting at MJD 59417.65584, is about 52 minutes after the estimated occurrence of the glitch at MJD 59417.6194(2). 
Since those three individual sub-observations contain enough pulses to make a SOM analysis we proceed to consider them individually independent. The results of those 6 SOM clustering studies are displayed in Fig.~\ref{fig:7-22A2}. 
\begin{table}
	\centering
	\caption{Date of selected observation with A2, the MJD at the beginning and end of the observation, the corresponding number of single pulses
	used for the pulse-by-pulse analysis of the 2021 Vela glitch, and the initial topocentric period, $P_\mathrm{obs}$. The estimated time of the glitch on July 22 is MJD 59417.6194(2).}
	\label{tab:A2}
	\begin{tabular}{l l l r r} 
		\hline
		Date & initial MJD & final MJD & \# pulses & $P_{\mathrm{obs}}$ [ms]\\
		\hline
		July 22 A22& 59417.65584 & 59417.68289 & 26131 & 89.414030\\
		July 22 A23& 59417.68317 & 59417.74006 & 54970 & 89.414042\\
		July 22 A24& 59417.74035 & 59417.76530 & 24117 & 89.414068\\
\hline
		July 20 A21& 59415.63988 & 59415.74171 & 98394 & 89.414230\\
		July 20 A22& 59415.74200 & 59415.77083 & 27853 & 89.414276\\
		\hline
	\end{tabular}
\end{table}

To supplement the information in Table~\ref{tab:A2} for the observations with A2 discussed here, we have that in total the observation time on July 22 is 2.65~h (divided into three observations) with a total SNR of 689, while on July 20, the (2) observations added up to 3.14~h with a total SNR of 814.
\begin{figure*}
   \centering 
\includegraphics[width=\textwidth]{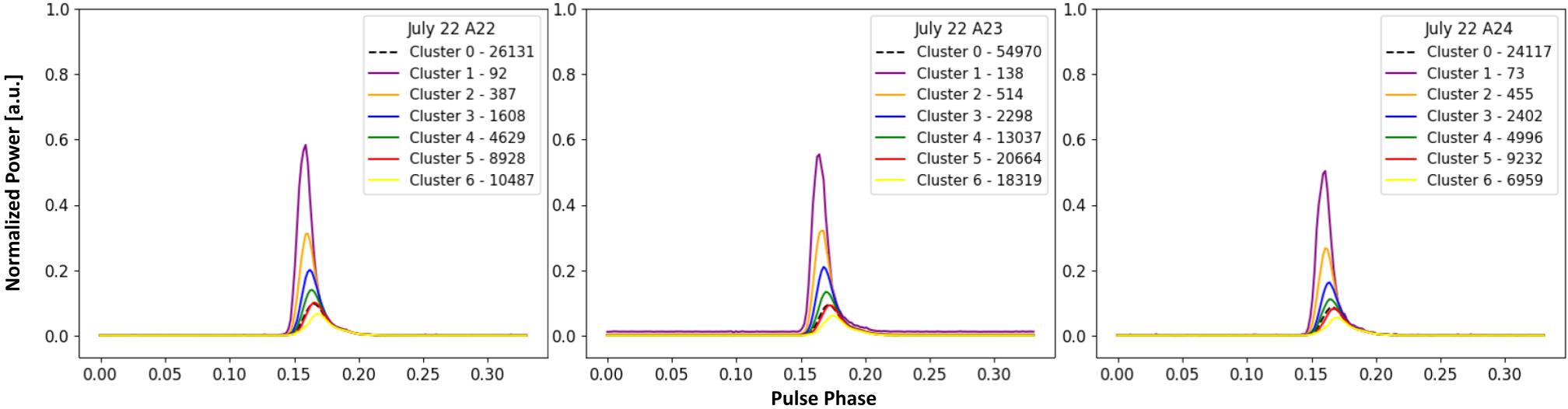}
     \caption{Mean cluster pulses for  2021 July 22 three successive observations (roughly 1–3.5 h after the glitch) with Antenna 2 for 6 SOM clusters with VAE reconstruction. 200 (out of total 611) phase bins were taken around the mean peak of each day to perform the single-pulse analysis on.}
  \label{fig:7-22A2}
 \end{figure*}
 
In Fig.~\ref{fig:7-22A2} we display the results for 6 SOM clustering for the two observations with A2 on the glitch day July 22 as described in Table~\ref{tab:A2}. We first observe that the right hand side of the mean cluster pulses seem to superpose and that the sequence of clusters, with increasing amplitude seem to appear earlier and earlier. The pulse width also shows a (weak) dependence on the cluster, being narrower for higher amplitude mean pulses. All these features, for the three observations covering from roughly 
1--3.5~h 
after this large glitch seem to be similar to those in between glitches, as we have observed in our previous analysis of the Vela pulses from January and March 2021 \citep{Lousto:2021dia}.

\subsubsection{On the 2021 July 20 observations}\label{sec:July20}

Given the unusual effects observed with A1 on July 20, we can cross check them against the corresponding A2 observations. The observations with A2 on July 20 have an interruption that split them into two observations as described in Table~\ref{tab:A2}. The first part of A2 observations start earlier than the A1 observation, and the second part of the A2 observation starts roughly about the last 30\% of the A1 observation, where the unusual effects are taking place according to our analysis in Figs. \ref{fig:Time4}-\ref{fig:Time9} and \ref{fig:7-20time}.

\begin{figure*}
   \centering 
     \includegraphics[width=0.65\linewidth]{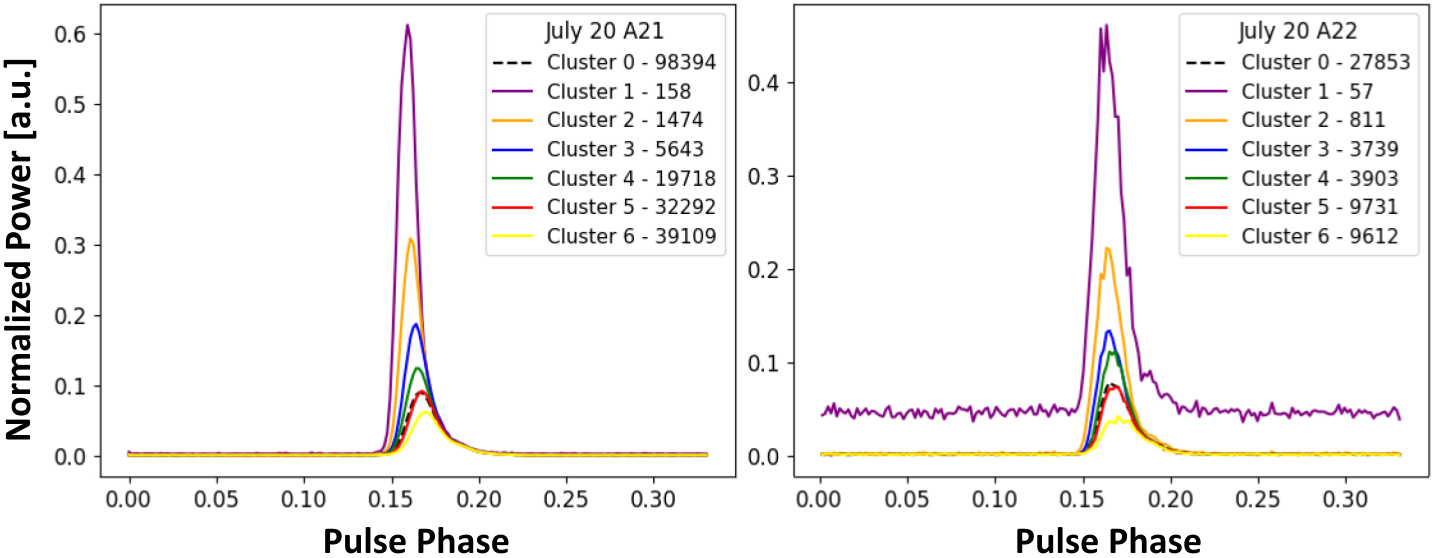}
     \caption{Mean cluster reconstruction for 2021, July 20 two observations on Antenna 2 for SOM 6 clustering with VAE pulses. 200 (out of total 611) phase bins were taken around the mean peak of each day to perform the single-pulse analysis on.} 
  \label{fig:7-20A2}
 \end{figure*}

In Fig.~\ref{fig:7-20A2} we display the Antenna 2 for 6 SOM clustering for the two observations on July 20 as described in table~\ref{tab:A2}.
We observe that the first observation shows the now standard pattern of mean pulses clusters ordered with increasing amplitude appearing earlier, being narrower, and a right "wing" superposition. On the other hand, the second observation shows a more shagged pulse structure, and the highest amplitude cluster displaying an increase in the baseline (noisy) emission. Since the second observation contains less pulses (27,853) than the first part (98,394) it would be expected some statistical noise, but on the other hand, we have just seen that the observation 1 and 3 with A2 of the glitch day, July 22, have less pulses but show smooth pulses structure. We can confirm now that there is a second part of the observations with A1 and A2 that display irregular features. We have not been able to discard them on the grounds of RFI or instrumental. The irregularities have different characteristics as seen with A1 or A2, but while A1 has a single polarization 112~MHz bandwidth, A2 has a two (circular) polarization sensitivity with 56~MHz of bandwidth. 

It is also important to point out here that the $DM$ is a crucial parameter in pulsar
timing. The Vela pulsar is known to have a constantly changing DM (see 
\cite{1985MNRAS.214P...5H,2013MNRAS.435.1610P,2021A&A...647A..25E}, for
example), however the time scales do not necessarily agree with the sudden change we found on the 2021 July 20 observation. We also checked that the variations in DM are below 0.2 pc cm$^{-3}$, which leads to offsets in pulse delays much smaller than the selected bin size.
Nevertheless this potential features require further study.

The observation of July 20 with A1 presents a distinctive feature with respect to the previous and posterior days to the glitch on July 22
as seen in Fig.~\ref{fig:All}.
Already at the level of 4 SOM cluster analysis a baseline displacement on the mean cluster pulses is observed.
The average pulse (cluster 0 labeled in this figure) does not show any atypical features, but introducing 9 SOM clusters reveals fluctuations in the baseline. After a more careful inspection presented in Appendix~\ref{sec:appendix2}, we conclude that these fluctuations are unrelated to the pulsar itself but instead due to local RFIs that were not properly removed. 

In Fig.~\ref{fig:7-20time} we display the detail of the number of pulses, labelled by an integer number index, (with the the side bar color map representing number density) versus time (given the ordered pulse identification number from the beginning of the observation). The 9 SOM cluster decomposition shows that while the first half of the observation ($\sim1.15$ hours) the distribution over the clusters follows a pattern similar to all the other days of observation, the second part of the observation ($\sim1.30$ hours) displays a clear shuffle of the number pulses from the medium/high amplitude clusters
towards the low amplitude ones. 

\begin{figure}
   \centering 
     \includegraphics[width=0.5\textwidth]{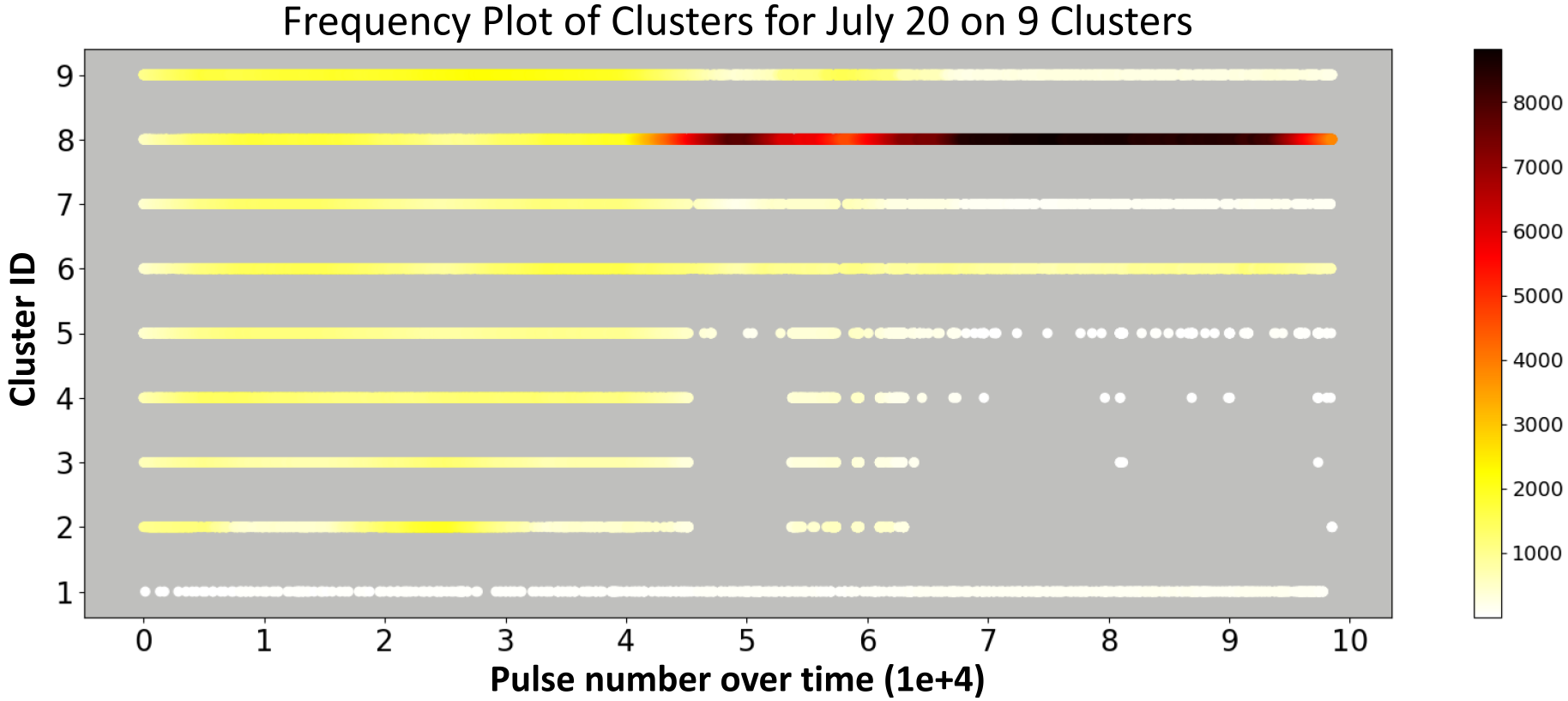}
      \caption{Time line distribution of number of pulses for the  2021, July 20 observations on Antenna 1 for 9 SOM clustering.}
  \label{fig:7-20time}
 \end{figure}

This suggest us to {\it artificially} split the A1 observation into those distinctive parts (roughly a 40\%/60\% split in time) and analyze them independently with our methods SOM clustering, as was done naturally with the two A2 observations of July 20. A second point is to instead of focusing the SOM clustering on zooming around the main pulse we will consider the whole period including the pulse. In this way the focus is rather on the complete baseline behavior we want to analyze in detail. The results are displayed in Fig.~\ref{fig:7-20} and are notably elucidating as we are able to
single out clusters with a sinusoidal behavior, covering roughly 9 periods during the Vela pulsar period of 11.18Hz leading to a period of very nearly $9\times11.18$Hz$\approx$100Hz. This is a strong evidence of the features in question are of a non astrophysical origin. In particular, the A/C power of the IAR being at 50Hz. We thus conclude that removing this feature, the pulse clustering on July 20 behaves qualitatively as the other previous days to the glitch.

\begin{figure}
   \centering 
     \includegraphics[width=0.5\textwidth]{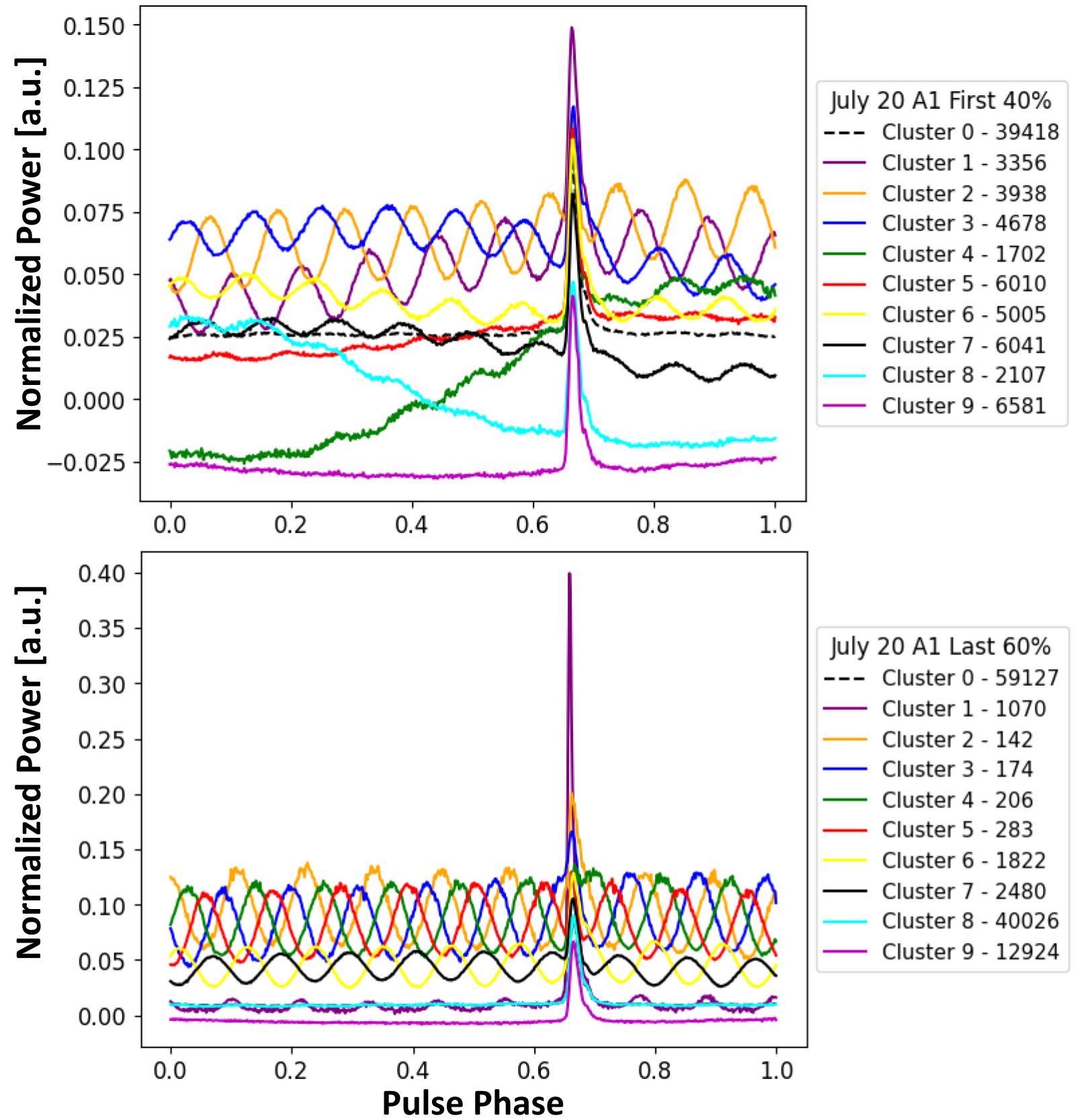}
     \caption{Mean cluster reconstruction for 2021 July 20 observation on Antenna 1 splitted 40\% / 60\% in time for SOM 9 clustering with VAE to exhibit the 100Hz interference in the whole phase range.}
  \label{fig:7-20}
 \end{figure}

\section{Conclusions and discussion}\label{sec:conclusions}

In this paper we have reported the first results of a southern glitching pulsars monitoring campaign at the Argentine Institute of Radioastronomy.
In 2019 we reported a large Vela $(\#21$ recorded) glitch \citep{atel_vela,Gancio2020} with a $(\Delta\nu_\mathrm{g}/\nu)_{2019}=2.7\times10^{-6}$. Here we report a detailed analysis of the latest $(\#22$ recorded) 2021 Vela glitch \citep{2021ATel14806....1S}, with comparable $(\Delta\nu_\mathrm{g}/\nu)_{2021}=1.2\times10^{-6}$,
providing an accurate description of the glitch characteristic epoch, jumps, and exponential recovery of 6.4 and 1 days times scales,
(See Table~\ref{tab:Vglitch} and Fig.~\ref{fig:Qtau}). The accuracy of our observations and procedures
allowed us to determine two mini-glitches (the smallest recorded so far) in PSR 1048$-$5832, $(\#8$ and $\#9$ recorded), with $(\Delta\nu_\mathrm{g}/\nu)_{2020}=8.9\times10^{-9}$ and $(\Delta\nu_\mathrm{g}/\nu)_{2021}=9.9\times10^{-9}$, respectively. These accuracy also allowed us to make pulse-by-pulse studies of Vela and use the machine learning techniques validated in \cite{Lousto:2021dia}.
Regarding the baseline features observed with A1 on July 20 in the pulse-by-pulse analysis, we have been able to identify its nature with a 100Hz interference that was not removed by the action of \texttt{RFIClean} and \texttt{rfifind} in tandem. This reveals the sensitivity of the pulse-by-pulse VAE/SOM analysis to extract features, in this case some sort of RFI, but eventually also others of astrophysical origin.



For the sake of completeness we mention here two recent glitches detected by our survey in PRS J0742$-$2822 and PSR J1740$-$3015 in sections \ref{sec:J0742} and \ref{sec:J1740} although they have not been studied yet in the same detail as 
PRS J1048$-$5832 and PSR J0835$-$4510 in sections \ref{sec:J0835} and \ref{sec:J1048}.

With the future improvements in IAR's antennas receivers, which include a combination of broader bandwidth and reduction of system temperature, it will be possible to study the dynamical spectra of single pulses for other pulsars of interest, such as PRS J1644$-$4559 and J0437-4715, the later not glitching but of interest to improve pulsar timing arrays data to detect a stochastic gravitational waves background. 


\section*{Acknowledgements}


We specially thank Yogesh Maan for numerous beneficial discussions about the best use of \texttt{RFIClean}, and Nils Andersson for discussion on the glitch theoretical modelling in regular pulsars / neutron stars. The authors also thank Cameron Knight for developing the early studies involving machine learning techniques, and Adolfo Simaz Bunzel in the observations reduction pipeline.
COL gratefully acknowledge the National Science Foundation (NSF) for financial support from Grants No.\ PHY-1912632, PHY-2207920  
and RIT-COS 2021-DRIG grant. JAC and FG are CONICET researchers. JAC is a Mar\'ia Zambrano researcher fellow funded by the European Union  -NextGenerationEU- (UJAR02MZ). This work received financial support from PICT-2017-2865 (ANPCyT) and PIP 0113 (CONICET). JAC and FG were also supported by grant PID2019-105510GB-C32/AEI/10.13039/501100011033 from the Agencia Estatal de Investigaci\'on of the Spanish Ministerio de Ciencia, Innovaci\'on y Universidades, and by Consejer\'{\i}a de Econom\'{\i}a, Innovaci\'on, Ciencia y Empleo of Junta de Andaluc\'{\i}a as research group FQM-322, as well as FEDER funds.


\section*{Data Availability}

Data generated by our calculations or observations are available from the corresponding authors upon reasonable request.



\bibliographystyle{mnras}
\bibliography{biblio,bibliografia,references} 




\appendix

\section{Tables of SOM Clustering}\label{sec:appendix}

Here we include the numerical information in tabular form about the 
clustering analysis summarized in Fig.~\ref{fig:All}. They include a 6 SOM clusters decomposition as representative for each of the days of observation.
We provide the number of pulses of each cluster \# pulses; peak location from the index of the maximum value in the pulse sequence;
peak height from the maximum value of the pulse sequence;
peak width done by first finding the maximum value of the sequence, then performing full-width half maximum of peak;
(library used for this: \url{https://docs.scipy.org/doc/scipy/reference/generated/scipy.signal.peak_widths.html});
for the peak skew we evaluated the Fisher-Pearson coefficient of skewness; 
(using the scipy for this computation \url{https://docs.scipy.org/doc/scipy/reference/generated/scipy.stats.skew.html});
The cluster \#0 corresponds to the total number of pulses in the observation and the successive clusters from \#1 to the \#6 SOM clustering are ordered accordingly to the highest peak amplitude of the mean pulse computed for each cluster and represented in Fig.~\ref{fig:All}.
We compute the peak location with respect to our grid of bins (here centered at around 100 for cluster \#0) and totaling 611 bins per period, giving us a time resolution of 146$\mu$s. We also provide a measure of the pulse width as given by the standard deviation $(\sigma)$ and its skewness, all with estimated $1-\sigma$ errors,
and finally MSE is the standard mean squared error $\sum_{i=1}^N(x_i-\bar{x})^2/N$, the average per-step mean squared reconstruction error over all sequences.
We observe a systematic tendency for the pulses' peaks to appear earlier the higher the amplitude as well as a reduction of its width and an increase of the skew (also observed in the previous work of \cite{Lousto:2021dia} analyzing  2021, Jan. 21, 24, 28 and March 29 observations), except for the especial case of the July 20 observations. 

\begin{table*}
	 \centering
	 \caption{SOM Clustering for July 19 with Antenna 1.}
	 \label{tab:[19b]}
	 \begin{tabular}{cclllll}
	 	 \hline
	 	 Cluster \# & \# Pulses & Peak Loc & Peak Height & Peak Width & Peak Skew & MSE \\ 
	 	 \hline 
	 	 0 & 143082 & $100.28 \pm 4.30$ & $13.31 \pm 9.81$ & $8.23 \pm 3.60$ & $3.39 \pm 0.76$ & $0.00004 \pm 0.00008$  \\ 
	 	 1 & 328 & $95.26 \pm 0.69$ & $129.84 \pm 43.14$ & $3.40 \pm 0.34$ & $6.84 \pm 0.57$ & $0.02966 \pm 0.13766$  \\ 
	 	 2 & 6973 & $97.14 \pm 0.99$ & $36.52 \pm 14.53$ & $3.61 \pm 0.55$ & $4.80 \pm 0.74$ & $0.00102 \pm 0.00243$  \\ 
	 	 3 & 55882 & $99.77 \pm 1.23$ & $14.90 \pm 4.39$ & $8.18 \pm 1.26$ & $3.65 \pm 0.38$ & $0.00011 \pm 0.00020$  \\ 
	 	 4 & 17810 & $100.22 \pm 11.30$ & $13.71 \pm 5.24$ & $12.63 \pm 5.98$ & $2.40 \pm 0.99$ & $0.00042 \pm 0.00071$  \\ 
	 	 5 & 49474 & $101.24 \pm 1.48$ & $8.59 \pm 1.76$ & $10.00 \pm 1.05$ & $3.31 \pm 0.40$ & $0.00012 \pm 0.00019$  \\ 
	 	 6 & 12615 & $100.75 \pm 1.56$ & $8.40 \pm 3.93$ & $10.07 \pm 1.57$ & $3.11 \pm 0.68$ & $0.00050 \pm 0.00084$  \\ 
	 	 \hline 
	 \end{tabular} 
\end{table*} 

\begin{table*}
	 \centering
	 \caption{SOM Clustering for  July 20 with Antenna 1.}
	 \label{tab:[20b]}
	 \begin{tabular}{cclllll}
	 	 \hline
	 	 Cluster \# & \# Pulses & Peak Loc & Peak Height & Peak Width & Peak Skew & MSE \\ 
	 	 \hline 
	 	 0 & 98545 & $99.66 \pm 15.13$ & $15.24 \pm 9.75$ & $15.00 \pm 9.45$ & $2.33 \pm 1.34$ & $0.00008 \pm 0.00019$  \\ 
	 	 1 & 1308 & $96.26 \pm 0.76$ & $65.79 \pm 29.64$ & $4.80 \pm 0.70$ & $4.83 \pm 0.86$ & $0.01173 \pm 0.07067$  \\ 
	 	 2 & 7542 & $98.14 \pm 26.17$ & $23.02 \pm 6.43$ & $24.39 \pm 10.35$ & $0.67 \pm 0.77$ & $0.00136 \pm 0.00239$  \\ 
	 	 3 & 10502 & $99.50 \pm 24.69$ & $17.97 \pm 5.06$ & $22.04 \pm 10.56$ & $0.83 \pm 0.85$ & $0.00092 \pm 0.00156$  \\ 
	 	 4 & 20454 & $100.51 \pm 17.77$ & $16.61 \pm 6.16$ & $12.76 \pm 5.91$ & $1.73 \pm 1.14$ & $0.00044 \pm 0.00079$  \\ 
	 	 5 & 47964 & $99.50 \pm 9.51$ & $13.36 \pm 5.98$ & $9.33 \pm 1.45$ & $2.99 \pm 0.95$ & $0.00015 \pm 0.00029$  \\ 
	 	 6 & 10775 & $100.35 \pm 3.36$ & $6.73 \pm 5.15$ & $9.57 \pm 1.38$ & $2.79 \pm 0.94$ & $0.00068 \pm 0.00118$  \\ 
	 	 \hline 
	 \end{tabular} 
\end{table*} 

\begin{table*}
	 \centering
	 \caption{SOM Clustering for  July 21 with Antenna 1.}
	 \label{tab:[21b]}
	 \begin{tabular}{cclllll}
	 	 \hline
	 	 Cluster \# & \# Pulses & Peak Loc & Peak Height & Peak Width & Peak Skew & MSE \\ 
	 	 \hline 
	 	 0 & 98948 & $99.95 \pm 1.59$ & $12.84 \pm 8.70$ & $9.27 \pm 1.62$ & $3.47 \pm 0.59$ & $0.00007 \pm 0.00012$  \\ 
	 	 1 & 501 & $95.42 \pm 0.73$ & $87.19 \pm 37.85$ & $3.60 \pm 0.43$ & $6.27 \pm 0.52$ & $0.01822 \pm 0.08315$  \\ 
	 	 2 & 6121 & $97.02 \pm 0.90$ & $29.68 \pm 10.70$ & $5.49 \pm 0.89$ & $4.48 \pm 0.51$ & $0.00122 \pm 0.00267$  \\ 
	 	 3 & 22475 & $98.89 \pm 0.97$ & $16.32 \pm 3.14$ & $7.42 \pm 0.83$ & $3.69 \pm 0.35$ & $0.00031 \pm 0.00055$  \\ 
	 	 4 & 21361 & $100.73 \pm 1.03$ & $10.81 \pm 1.84$ & $9.84 \pm 0.97$ & $3.06 \pm 0.56$ & $0.00032 \pm 0.00051$  \\ 
	 	 5 & 41055 & $100.60 \pm 1.25$ & $9.32 \pm 2.10$ & $9.69 \pm 1.00$ & $3.45 \pm 0.34$ & $0.00016 \pm 0.00025$  \\ 
	 	 6 & 7435 & $100.04 \pm 1.53$ & $8.68 \pm 3.68$ & $9.60 \pm 1.32$ & $3.05 \pm 0.60$ & $0.00092 \pm 0.00150$  \\ 
	 	 \hline 
	 \end{tabular} 
\end{table*} 

\begin{table*}
	 \centering
	 \caption{SOM Clustering for July 22 with Antenna A2, Observation 1.}
	 \label{tab:[22b1]}
	 \begin{tabular}{cclllll}
	 	 \hline
	 	 Cluster \# & \# Pulses & Peak Loc & Peak Height & Peak Width & Peak Skew & MSE \\ 
	 	 \hline 
	 	 0 & 26131 & $100.15 \pm 1.70$ & $16.35 \pm 8.57$ & $9.26 \pm 0.67$ & $3.54 \pm 0.25$ & $0.00024 \pm 0.00040$  \\ 
	 	 1 & 92 & $95.65 \pm 0.50$ & $88.78 \pm 19.49$ & $7.70 \pm 0.29$ & $4.41 \pm 0.10$ & $0.12615 \pm 0.58499$  \\ 
	 	 2 & 387 & $96.66 \pm 0.91$ & $48.20 \pm 7.94$ & $8.39 \pm 0.18$ & $4.12 \pm 0.13$ & $0.01874 \pm 0.03667$  \\ 
	 	 3 & 1609 & $97.92 \pm 0.97$ & $31.15 \pm 3.70$ & $8.89 \pm 0.28$ & $3.88 \pm 0.12$ & $0.00412 \pm 0.00667$  \\ 
	 	 4 & 4627 & $98.68 \pm 1.06$ & $21.75 \pm 2.76$ & $9.51 \pm 0.56$ & $3.70 \pm 0.16$ & $0.00140 \pm 0.00220$  \\ 
	 	 5 & 8927 & $100.00 \pm 1.11$ & $15.68 \pm 2.21$ & $9.87 \pm 0.43$ & $3.56 \pm 0.17$ & $0.00071 \pm 0.00109$  \\ 
	 	 6 & 10489 & $101.43 \pm 1.20$ & $10.45 \pm 1.61$ & $9.91 \pm 0.71$ & $3.37 \pm 0.22$ & $0.00060 \pm 0.00094$  \\ 
	 	 \hline 
	 \end{tabular} 
\end{table*} 

\begin{table*}
	 \centering
	 \caption{SOM Clustering for July 22 with Antenna A23.}
	 \label{tab:[22b2]}
	 \begin{tabular}{cclllll}
	 	 \hline
	 	 Cluster \# & \# Pulses & Peak Loc & Peak Height & Peak Width & Peak Skew & MSE \\ 
	 	 \hline 
	 	 0 & 54970 & $99.76 \pm 1.69$ & $15.57 \pm 7.79$ & $9.93 \pm 0.68$ & $3.56 \pm 0.20$ & $0.00014 \pm 0.00204$  \\ 
	 	 1 & 138 & $94.76 \pm 1.08$ & $85.77 \pm 18.33$ & $7.98 \pm 1.67$ & $4.27 \pm 1.00$ & $0.06816 \pm 0.18188$  \\ 
	 	 2 & 515 & $96.17 \pm 0.97$ & $50.06 \pm 7.30$ & $8.21 \pm 0.27$ & $4.15 \pm 0.18$ & $0.01428 \pm 0.02705$  \\ 
	 	 3 & 2296 & $97.17 \pm 0.82$ & $32.33 \pm 3.86$ & $8.83 \pm 0.34$ & $3.93 \pm 0.10$ & $0.00299 \pm 0.00585$  \\ 
	 	 4 & 13034 & $98.23 \pm 1.07$ & $20.82 \pm 3.07$ & $9.58 \pm 0.29$ & $3.71 \pm 0.11$ & $0.00050 \pm 0.00080$  \\ 
	 	 5 & 20667 & $99.82 \pm 1.04$ & $14.41 \pm 1.95$ & $10.23 \pm 0.35$ & $3.56 \pm 0.11$ & $0.00030 \pm 0.00048$  \\ 
	 	 6 & 18320 & $101.24 \pm 1.11$ & $9.55 \pm 1.45$ & $9.88 \pm 0.52$ & $3.40 \pm 0.14$ & $0.00058 \pm 0.01059$  \\ 
	 	 \hline 
	 \end{tabular} 
\end{table*} 

\begin{table*}
	 \centering
	 \caption{SOM Clustering for July 22 with Antenna A2, Observation 3.}
	 \label{tab:[22b3]}
	 \begin{tabular}{cclllll}
	 	 \hline
	 	 Cluster \# & \# Pulses & Peak Loc & Peak Height & Peak Width & Peak Skew & MSE \\ 
	 	 \hline 
	 	 0 & 24117 & $100.66 \pm 1.72$ & $14.58 \pm 7.34$ & $9.71 \pm 0.68$ & $3.52 \pm 0.24$ & $0.00026 \pm 0.00044$  \\ 
	 	 1 & 73 & $96.73 \pm 0.63$ & $76.18 \pm 13.84$ & $7.74 \pm 0.36$ & $4.39 \pm 0.10$ & $0.14697 \pm 0.71118$  \\ 
	 	 2 & 455 & $97.49 \pm 0.70$ & $41.17 \pm 6.94$ & $8.12 \pm 0.32$ & $4.09 \pm 0.11$ & $0.01556 \pm 0.03238$  \\ 
	 	 3 & 2402 & $98.68 \pm 0.94$ & $25.22 \pm 3.33$ & $9.29 \pm 0.37$ & $3.82 \pm 0.09$ & $0.00268 \pm 0.00427$  \\ 
	 	 4 & 4997 & $99.29 \pm 1.02$ & $17.34 \pm 2.32$ & $9.96 \pm 0.42$ & $3.60 \pm 0.14$ & $0.00127 \pm 0.00197$  \\ 
	 	 5 & 9232 & $100.96 \pm 1.16$ & $12.83 \pm 1.95$ & $9.87 \pm 0.30$ & $3.51 \pm 0.17$ & $0.00067 \pm 0.00103$  \\ 
	 	 6 & 6958 & $102.19 \pm 1.16$ & $8.85 \pm 1.10$ & $10.46 \pm 1.31$ & $3.33 \pm 0.19$ & $0.00091 \pm 0.00154$  \\ 
	 	 \hline 
	 \end{tabular} 
\end{table*} 

\begin{table*}
	 \centering
	 \caption{SOM Clustering for July 23 with Antenna 1.}
	 \label{tab:[23b]}
	 \begin{tabular}{cclllll}
	 	 \hline
	 	 Cluster \# & \# Pulses & Peak Loc & Peak Height & Peak Width & Peak Skew & MSE \\ 
	 	 \hline 
	 	 0 & 88740 & $100.03 \pm 1.55$ & $12.45 \pm 9.27$ & $8.13 \pm 1.81$ & $3.45 \pm 0.63$ & $0.00007 \pm 0.00013$  \\ 
	 	 1 & 133 & $95.05 \pm 0.68$ & $141.15 \pm 40.55$ & $3.25 \pm 0.29$ & $6.94 \pm 0.40$ & $0.07106 \pm 0.23160$  \\ 
	 	 2 & 1610 & $96.29 \pm 0.82$ & $50.89 \pm 20.57$ & $3.96 \pm 0.67$ & $5.46 \pm 0.66$ & $0.00484 \pm 0.01265$  \\ 
	 	 3 & 17975 & $98.58 \pm 1.19$ & $17.95 \pm 5.35$ & $7.67 \pm 1.38$ & $3.84 \pm 0.47$ & $0.00038 \pm 0.00071$  \\ 
	 	 4 & 6501 & $100.18 \pm 1.40$ & $13.78 \pm 4.03$ & $8.79 \pm 1.37$ & $2.78 \pm 0.72$ & $0.00107 \pm 0.00175$  \\ 
	 	 5 & 50900 & $100.59 \pm 1.18$ & $9.70 \pm 2.15$ & $8.60 \pm 0.85$ & $3.40 \pm 0.39$ & $0.00012 \pm 0.00020$  \\ 
	 	 6 & 11621 & $100.33 \pm 1.35$ & $8.47 \pm 3.66$ & $9.72 \pm 1.46$ & $3.11 \pm 0.62$ & $0.00056 \pm 0.00091$  \\ 
	 	 \hline 
	 \end{tabular} 
\end{table*} 

\begin{table*}
	 \centering
	 \caption{SOM Clustering for July 24 with Antenna 1.}
	 \label{tab:[24b]}
	 \begin{tabular}{cclllll}
	 	 \hline
	 	 Cluster \# & \# Pulses & Peak Loc & Peak Height & Peak Width & Peak Skew & MSE \\ 
	 	 \hline 
	 	 0 & 13401 & $100.42 \pm 1.73$ & $7.82 \pm 3.17$ & $9.74 \pm 1.27$ & $3.23 \pm 0.34$ & $0.00052 \pm 0.00256$  \\ 
	 	 1 & 424 & $99.73 \pm 1.68$ & $15.01 \pm 4.92$ & $8.91 \pm 0.87$ & $3.33 \pm 0.30$ & $0.02562 \pm 0.42168$  \\ 
	 	 2 & 1223 & $98.48 \pm 1.22$ & $12.94 \pm 2.46$ & $9.27 \pm 0.76$ & $3.60 \pm 0.19$ & $0.00624 \pm 0.02066$  \\ 
	 	 3 & 1622 & $100.71 \pm 1.48$ & $9.66 \pm 1.44$ & $9.38 \pm 1.16$ & $3.24 \pm 0.24$ & $0.00429 \pm 0.00654$  \\ 
	 	 4 & 3959 & $100.39 \pm 1.61$ & $8.06 \pm 1.49$ & $9.77 \pm 1.32$ & $3.33 \pm 0.28$ & $0.00168 \pm 0.00266$  \\ 
	 	 5 & 3292 & $100.46 \pm 1.77$ & $6.57 \pm 1.53$ & $9.66 \pm 1.68$ & $3.20 \pm 0.33$ & $0.00203 \pm 0.00356$  \\ 
	 	 6 & 2881 & $101.18 \pm 1.48$ & $4.64 \pm 1.05$ & $10.61 \pm 1.96$ & $2.92 \pm 0.26$ & $0.00237 \pm 0.00374$  \\ 
	 	 \hline 
	 \end{tabular} 
\end{table*}

\section{VAE reconstruction and SOM Clustering for July 20 observation with A1}\label{sec:appendix2}

In order to show that what we observe with the clusters baseline is not an artifact of the VAE pulse reconstruction method,
in Fig.~\ref{fig:7-20pulses} we display some selected {\it individual} raw pulses belonging to the 4 SOM clusters versus their corresponding reconstructions showing the actual baseline fluctuations over the full period range.

 \begin{figure*}
    \centering 
      \includegraphics[width=0.65\textwidth]{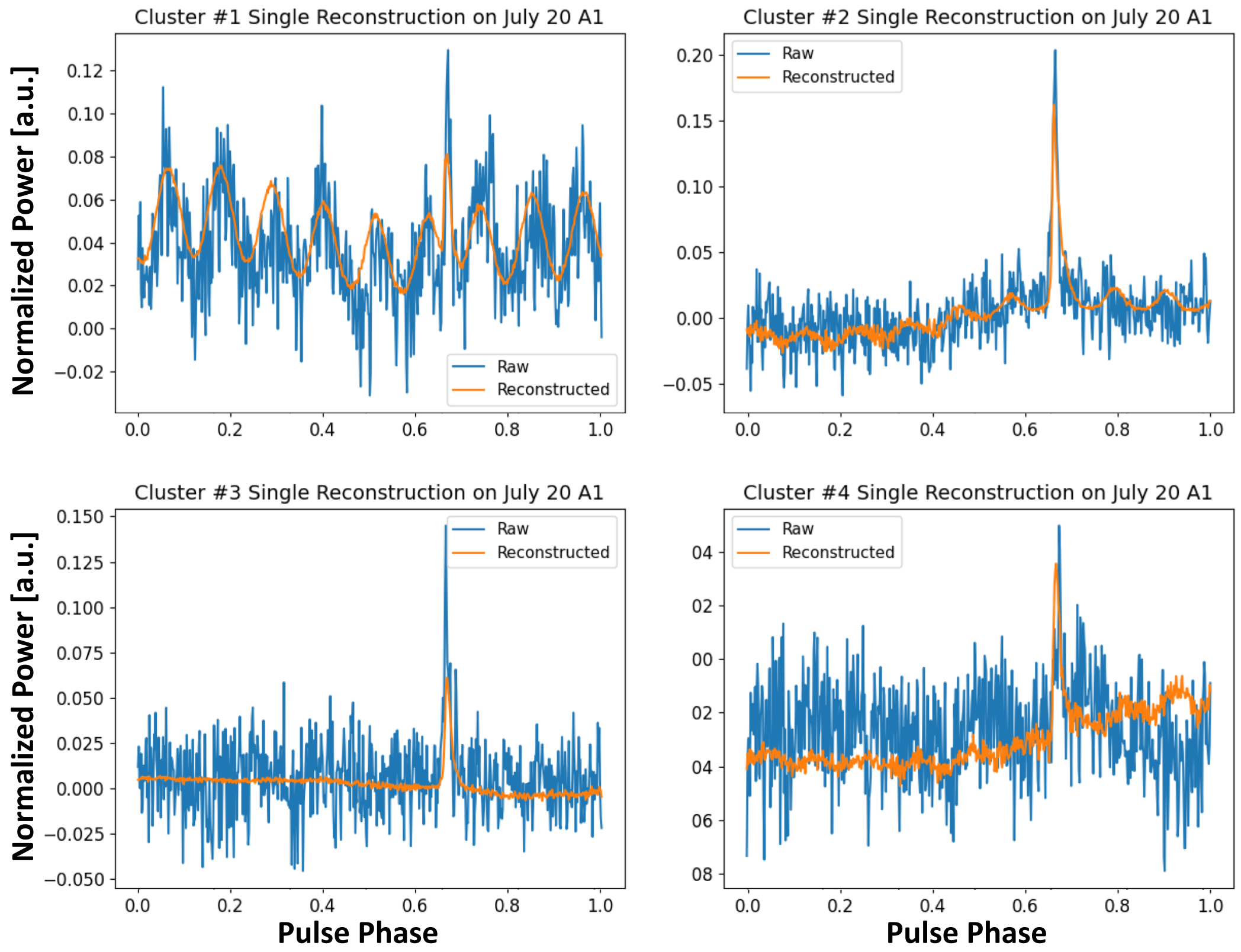}
      \caption{Sample of VAE pulse reconstruction for  2021, July 20 observations with A1 for 4 SOM clustering. }
   \label{fig:7-20pulses}
  \end{figure*}

The corresponding 4 clusters with SOM are displayed in Fig.~\ref{fig:7-20SOM} with the 100Hz baseline RFI.
\begin{figure}
   \centering 
     \includegraphics[width=0.34\textwidth]{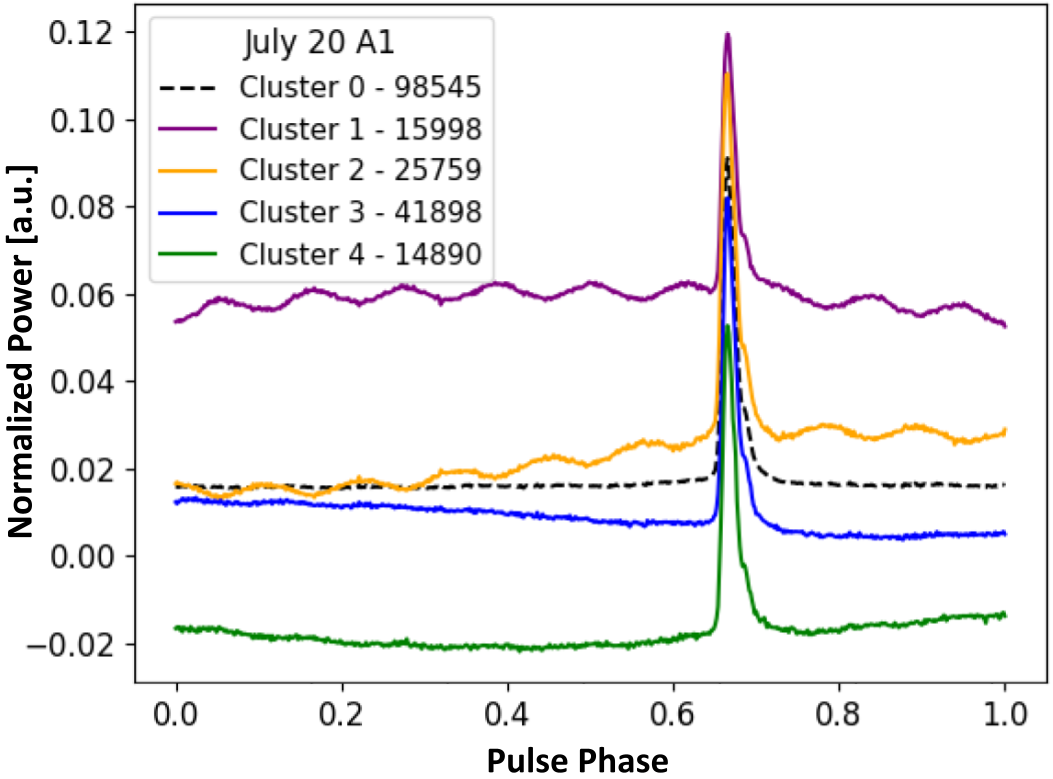}
     \caption{SOM 4 clustering for 2021, July 20 observations with A1 over the full period range.}
  \label{fig:7-20SOM}
 \end{figure}

\bsp	
\label{lastpage}
\end{document}